\documentclass[10pt,twocolumn,letterpaper]{article}

\usepackage{cvpr}              % To produce the CAMERA-READY version
\usepackage[accsupp]{axessibility}
\usepackage{graphicx}
\usepackage{amsmath}
\usepackage{amssymb}
\usepackage{booktabs}
\usepackage{multicol}
\usepackage{multirow}
\usepackage{subcaption}
\usepackage{adjustbox}

\usepackage[pagebackref,breaklinks,colorlinks]{hyperref}

\usepackage[capitalize]{cleveref}
\crefname{section}{Sec.}{Secs.}
\Crefname{section}{Section}{Sections}
\Crefname{table}{Table}{Tables}
\crefname{table}{Tab.}{Tabs.}

\def\cvprPaperID{39} % *** Enter the CVPR Paper ID here
\def\confName{CVPR}
\def\confYear{2023}

\begin{document}

%%%%%%%%% TITLE - PLEASE UPDATE
\title{OPDN: Omnidirectional Position-aware Deformable Network for Omnidirectional Image Super-Resolution}

% \author{First Author\\
% Institution1\\
% Institution1 address\\
% {\tt\small firstauthor@i1.org}
% % For a paper whose authors are all at the same institution,
% % omit the following lines up until the closing ``}''.
% % Additional authors and addresses can be added with ``\and'',
% % just like the second author.
% % To save space, use either the email address or home page, not both
% \and
% Second Author\\
% Institution2\\
% First line of institution2 address\\
% {\tt\small secondauthor@i2.org}
% }
% \renewcommand\footnotemark{}

\author{
Xiaopeng Sun$^{*1}$ \quad Weiqi Li\thanks{Equal contribution. Weiqi Li is an intern in MMLab, ByteDance. }\hspace{4pt}$^{1,2}$ \quad Zhenyu Zhang$^{1, 2}$ \quad Qiufang Ma$^{1}$ \quad Xuhan Sheng$^{2}$ \quad Ming Cheng$^{1}$ \\
\quad Haoyu Ma$^{1}$ \quad Shijie Zhao\thanks{Corresponding author. (e-mail: zhaoshijie.0526@bytedance.com) }\hspace{4pt}$^{1}$ \quad Jian Zhang$^{2}$ \quad Junlin Li$^{1}$ \quad Li Zhang$^{1}$\\
$^{1}$ByteDance Inc,\quad $^{2}$Peking University Shenzhen Graduate School\\
{\tt\small sunxiaopeng.01@bytedance.com, liweiqi@stu.pku.edu.cn}\\
}
\maketitle

%%%%%%%%% ABSTRACT
\begin{abstract}
   360° omnidirectional images have gained research attention due to their immersive and interactive experience, particularly in AR/VR applications. However, they suffer from lower angular resolution due to being captured by fisheye lenses with the same sensor size for capturing planar images. To solve the above issues, we propose a two-stage framework for 360° omnidirectional image super-resolution.
   The first stage employs two branches: model A, which incorporates omnidirectional position-aware deformable blocks (OPDB) and Fourier upsampling, and model B, which adds a spatial frequency fusion module (SFF) to model A. Model A aims to enhance the feature extraction ability of 360° image  positional information, while Model B further focuses on the high-frequency information of 360° images. The second stage performs same-resolution enhancement based on the structure of model A with a pixel unshuffle operation. In addition, we collected data from YouTube to  improve the fitting ability of the transformer, and created pseudo low-resolution images using a degradation network. Our proposed method achieves superior performance and wins the NTIRE 2023 challenge of 360° omnidirectional image super-resolution.
\end{abstract}

%%%%%%%%% BODY TEXT
\section{Introduction}
\label{sec:intro}

\begin{figure}[t!]
    \centering
    \begin{tabular}{cc}
         \includegraphics[width=0.45\linewidth]{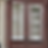}& \includegraphics[width=0.45\linewidth]{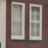} \\
         Bicubic & OSRT~\cite{yu2023osrt} \\
         \includegraphics[width=0.45\linewidth]{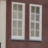}& \includegraphics[width=0.45\linewidth]{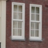}\\
         OPDN (Ours) & GroundTruth \\
    \end{tabular}
    \caption{Visual comparisons of $\times 4$ SR results on one image from Flickr360 validation set. We fine-tuned the OSRT~\cite{yu2023osrt} model on Flickr360 training set for fair comparison.}
    \label{fig:my_label}
\end{figure}

\begin{figure*}
    \centering
    % \vspace{0.5cm}
        \begin{tabular}{cccc}
    OSRT~\cite{yu2023osrt} & & & OPDN (ours)\\
    % \multicolumn{2}{c}{OSRT} & \multicolumn{2}{c}{OPDN (ours)}\\
    \includegraphics[width=0.3\textwidth]{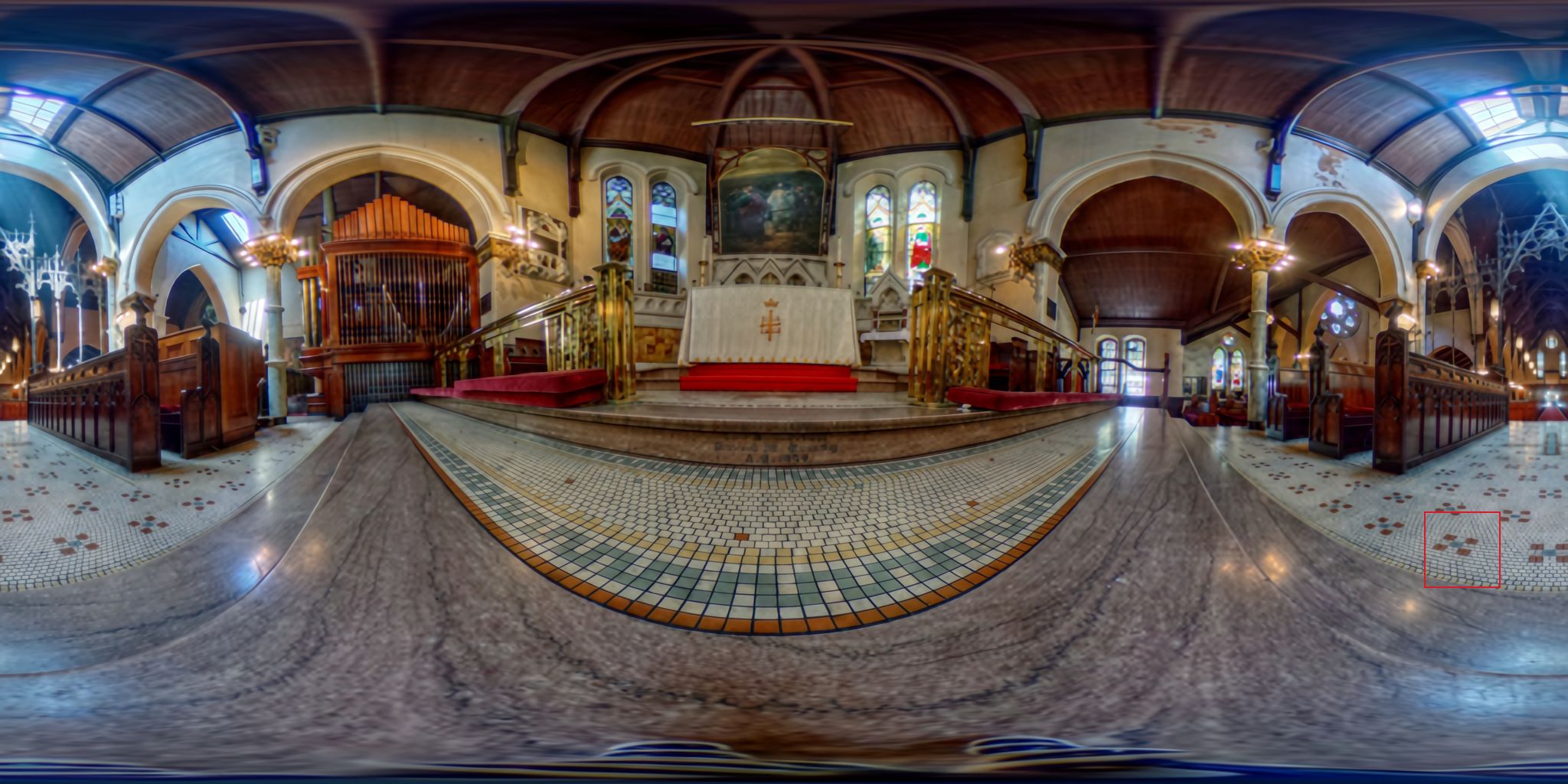}&
    \includegraphics[width=0.15\textwidth]{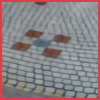}&
    \includegraphics[width=0.15\textwidth]{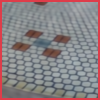}&
    \includegraphics[width=0.3\textwidth]{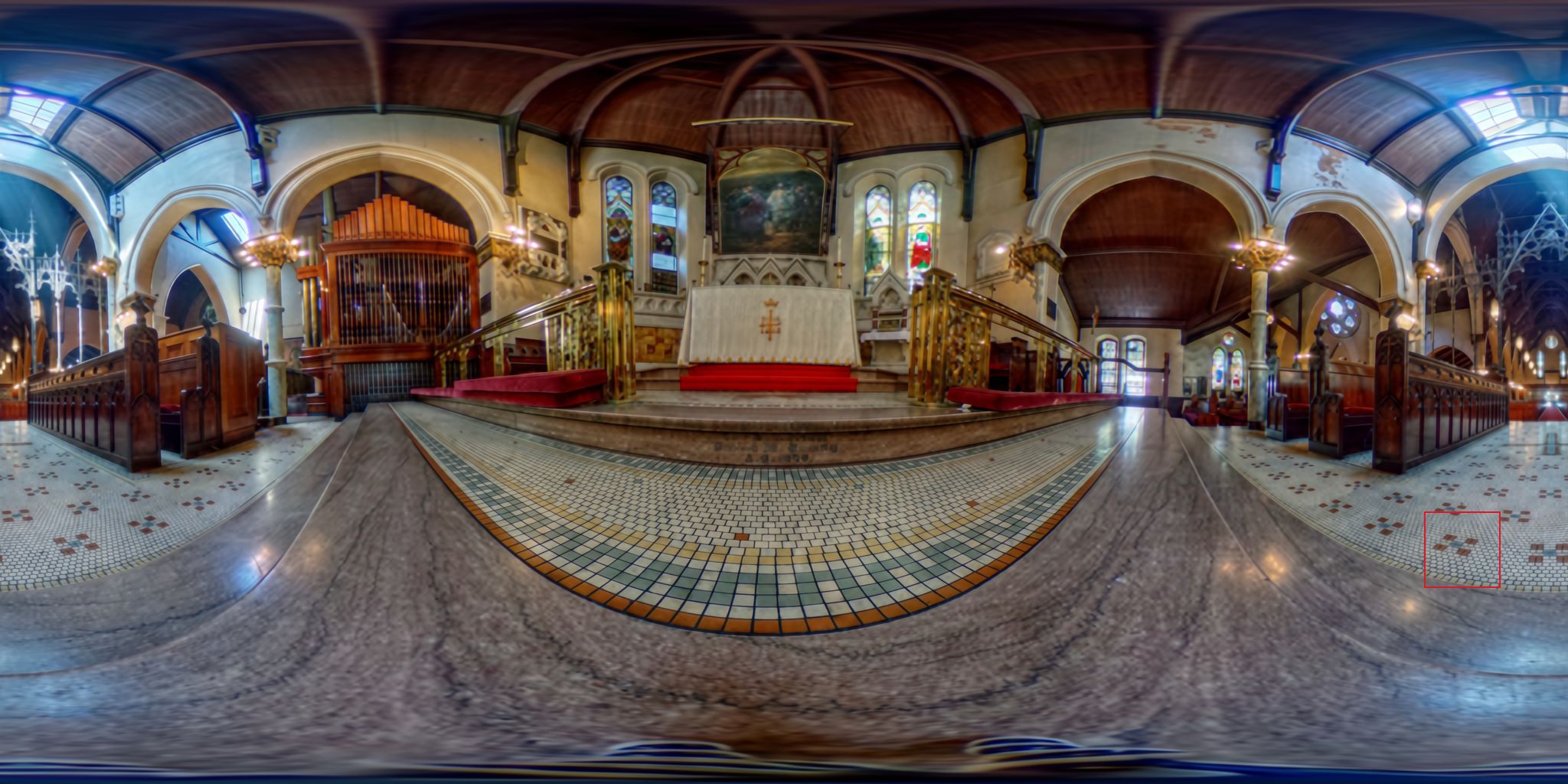}\\
    \includegraphics[width=0.3\textwidth]{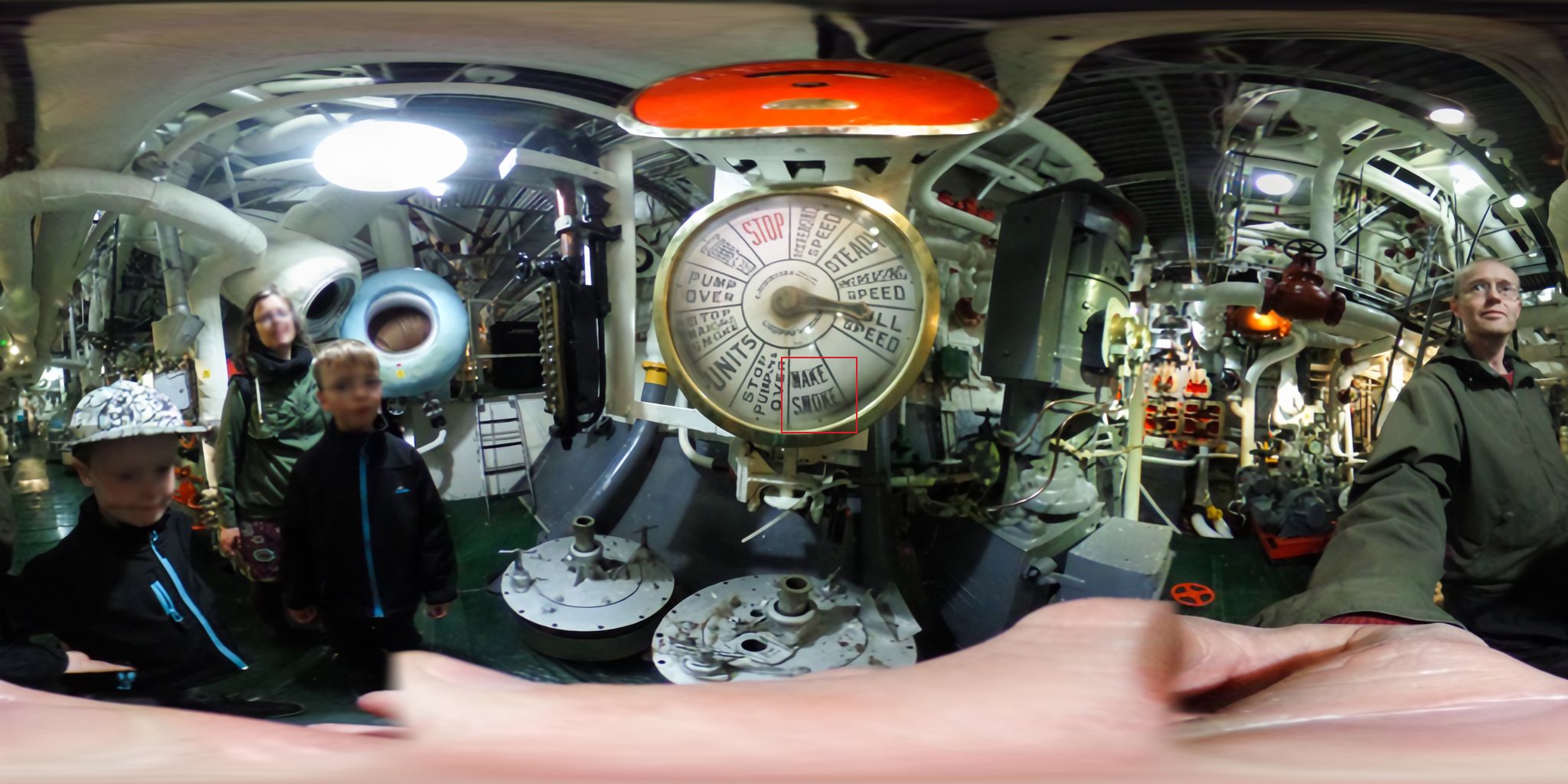}&
    \includegraphics[width=0.15\textwidth]{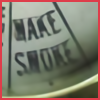}&
    \includegraphics[width=0.15\textwidth]{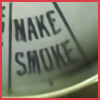}&
    \includegraphics[width=0.3\textwidth]{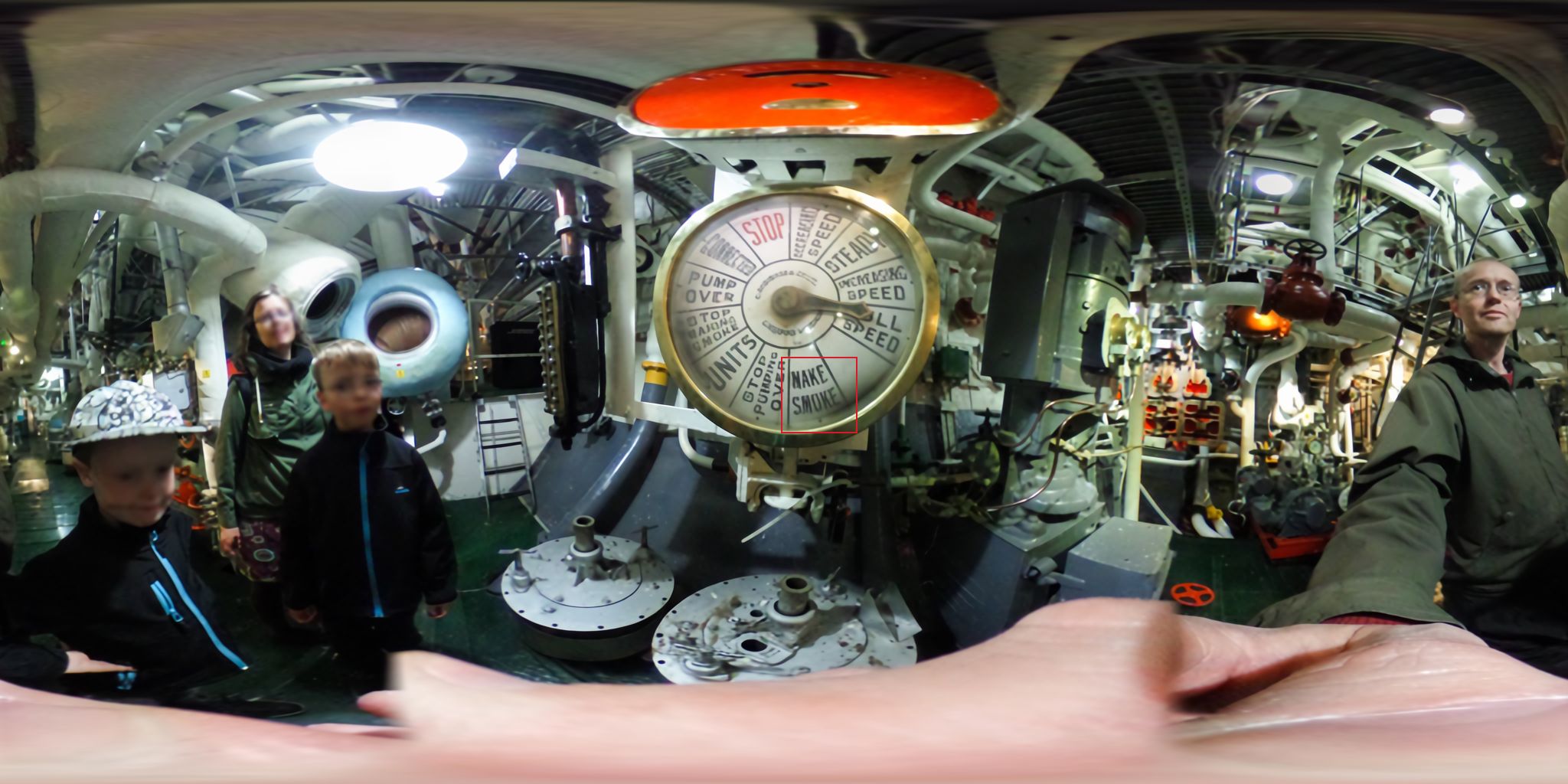}\\
    \includegraphics[width=0.3\textwidth]{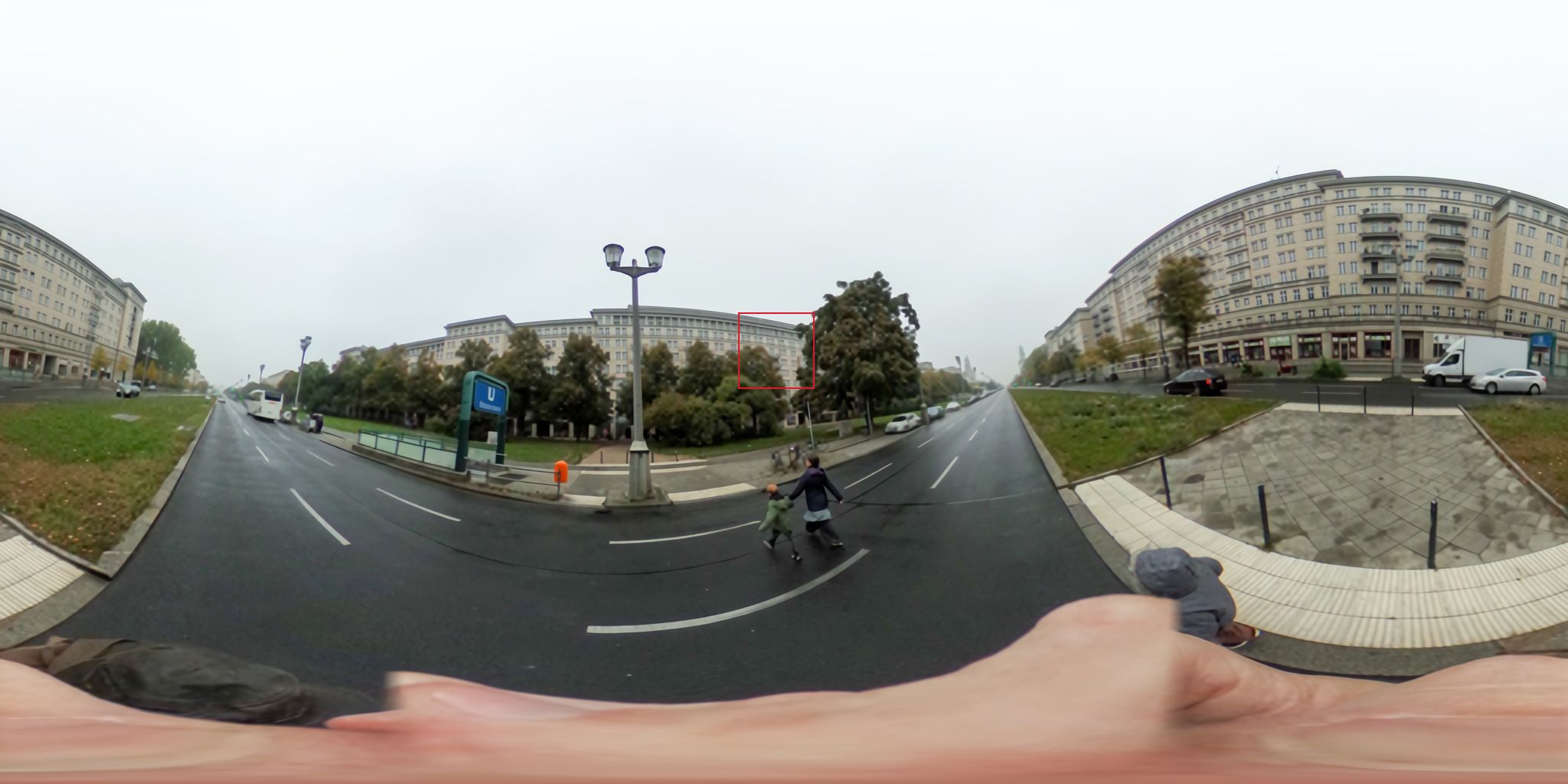}&
    \includegraphics[width=0.15\textwidth]{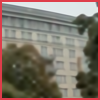}&
    \includegraphics[width=0.15\textwidth]{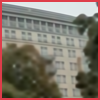}&
    \includegraphics[width=0.3\textwidth]{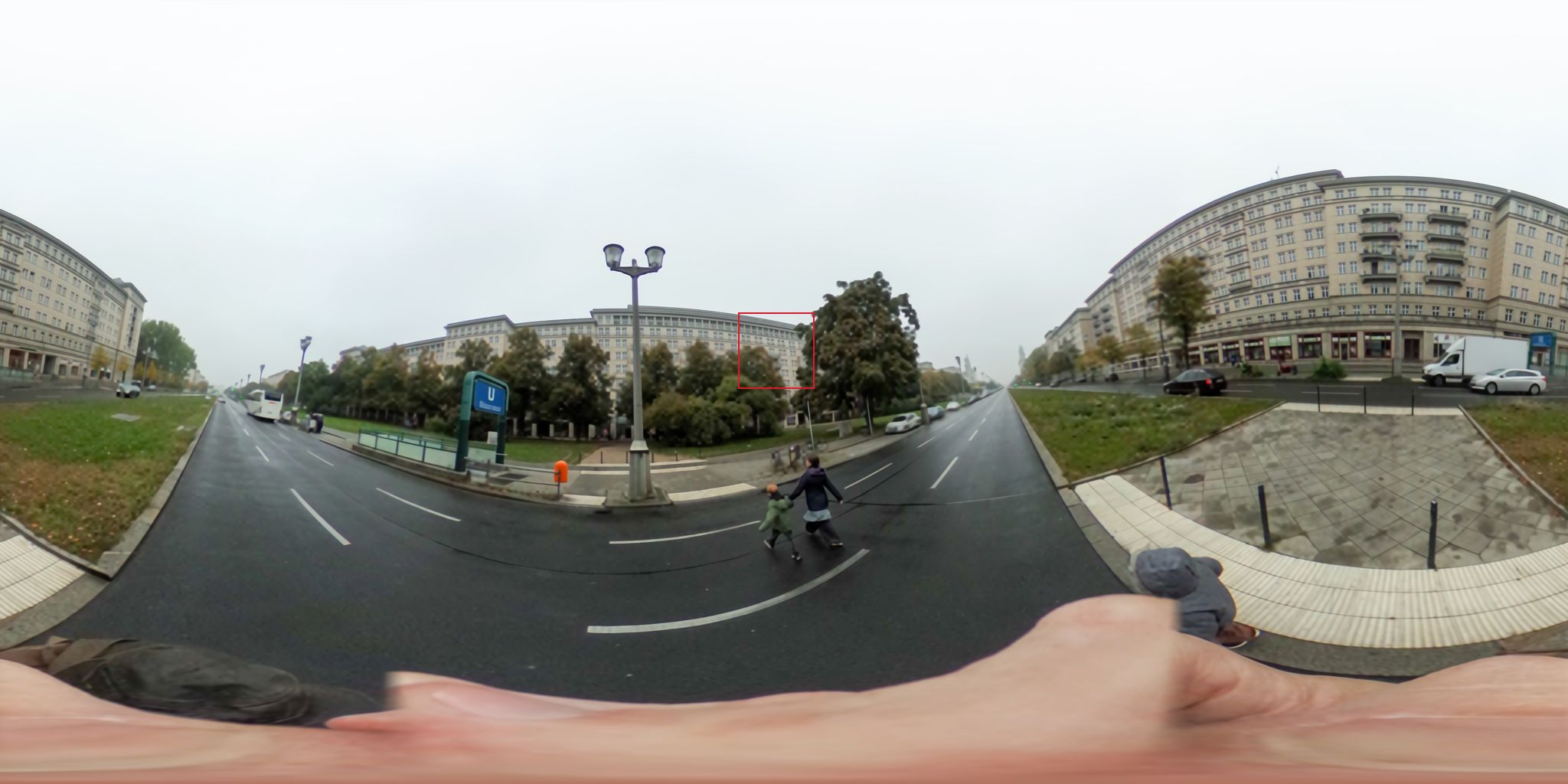}\\
    \includegraphics[width=0.3\textwidth]{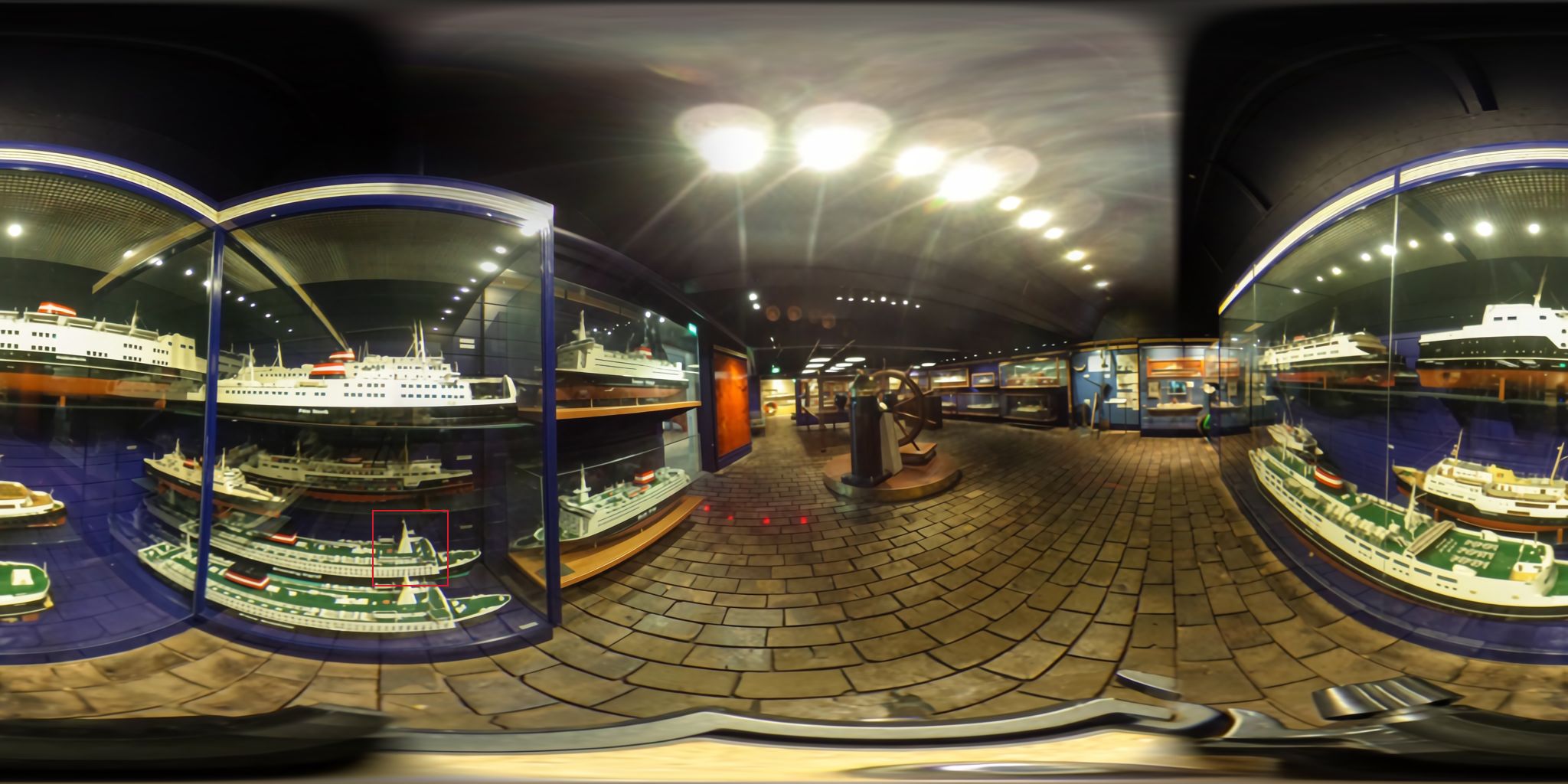}&
    \includegraphics[width=0.15\textwidth]{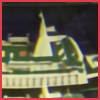}&
    \includegraphics[width=0.15\textwidth]{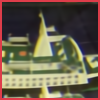}&
    \includegraphics[width=0.3\textwidth]{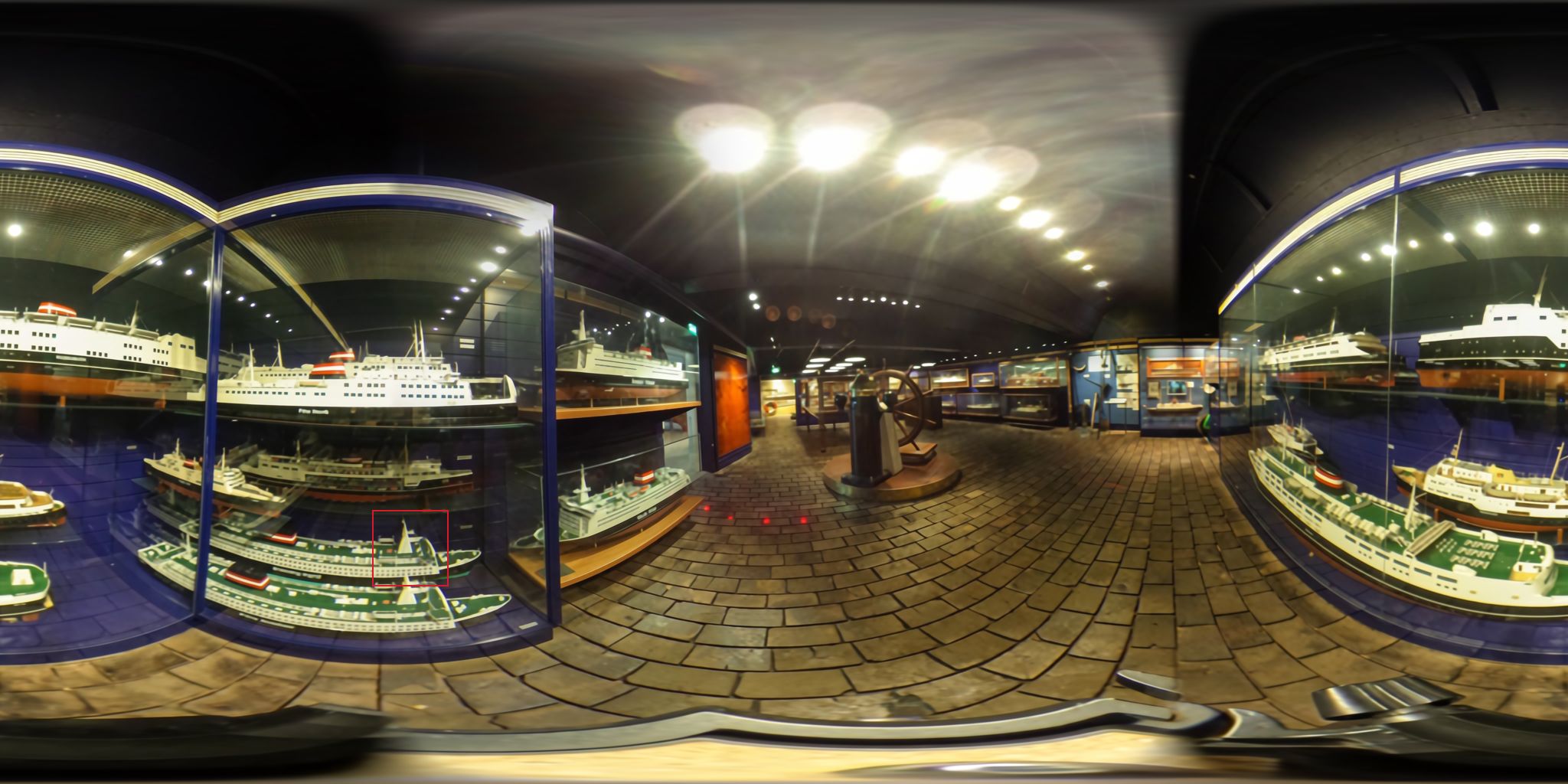}\\
\end{tabular}
    \vspace{0.1cm}
    \vskip -0.2cm
    \captionof{figure}{Visual comparisons of four images from Flickr360 validation set.}
    \label{fig:visual_val}
    \vspace{-0.3cm}
\end{figure*}
% With the popularity of AR/VR applications, 360° images, also known as omnidirectional images or panoramic images, have received much research interest in the computer vision community for its immersive and interactive propriety. Even the viewport range of 360° images is 360 $\times$ 180°, only a narrow field-of-view (FOV) is visible through head-mounted displays (HMDs). Therefore, an extremely high resolution, \textit{e.g.} 4K $\times$ 8K~\cite{ai2022deep}, is required to ensure a small viewport with sufficient details. Therefore, it is imperative to super-resolve the low resolution (LR) 360° image to provide a high visual quality.
With the rising popularity of AR/VR applications, 360° images, also known as omnidirectional or panoramic images, have garnered significant research interest in the computer vision community due to their immersive and interactive properties. Although the viewport range of 360° images is 360 × 180°, only a narrow field-of-view (FOV) is visible through head-mounted displays (HMDs). Consequently, extremely high resolutions, such as 4K × 8K~\cite{ai2022deep}, are required to ensure a small viewport with sufficient details. Thus, it is crucial to super-resolve low-resolution (LR) 360° images to provide high visual quality.

% Recently, deep learning (DL) has brought considerable success to the single image super-resolution (SISR). After the first introduction of DL in~\cite{dong2015image}, further works boost SR performance by CNNs~\cite{lim2017enhanced, shi2016real, zhang2018image, sun2019distilling, mei2021image, li2022d3c2, mou2022metric, mou2022transcl}, generative adversarial networks (GANs)~\cite{ledig2017photo, wang2018esrgan, wang2021real, zhang2019ranksrgan} and Vision Transformers (ViTs)~\cite{chen2021pre, chen2205activating, liang2021swinir, li2021efficient}. In particular, SwinIR~\cite{liang2021swinir} achieved impressive performance based on the Swin Transformer architecture, with a shifted window mechanism to model long-range dependencies. HAT~\cite{chen2205activating} further expanded the scale of the model, with a novel overlapping cross-attention to activate more pixels in the transformer, thus achieving state-of-the-art (SOTA) performance in the SISR task. However, these methods cannot be directly applied to 360° images, for the nonuniform pixel density and texture complexity across latitudes.
Recently, deep learning (DL) has significantly contributed to the success of single-image super-resolution (SISR). Following the initial introduction of DL in~\cite{dong2015image}, subsequent studies have enhanced SR performance using convolutional neural networks (CNNs)~\cite{lim2017enhanced, shi2016real, zhang2018image, sun2019distilling, mei2021image, li2022d3c2, mou2022metric, mou2022transcl}, generative adversarial networks (GANs)~\cite{ledig2017photo, wang2018esrgan, wang2021real, zhang2019ranksrgan, wang2023gpsr, hu2023DEARGAN}, Vision Transformers (ViTs)~\cite{chen2021pre, chen2205activating, liang2021swinir, li2021efficient} and diffusion models~\cite{wang2023ddnm}. Notably, SwinIR\cite{liang2021swinir} achieved remarkable performance utilizing the Swin Transformer architecture, incorporating a shifted window mechanism for modeling long-range dependencies. HAT~\cite{chen2205activating} further expanded the model's scale, introducing a novel overlapping cross-attention mechanism to activate more pixels in the transformer, thereby achieving state-of-the-art (SOTA) performance in the SISR task. However, these methods cannot be directly applied to 360° images due to nonuniform pixel density and texture complexity across latitudes.

% Consequently, some attempts have been made to address the 360° SR problems. LAU-Net~\cite{deng2021lau} splits the whole ERP image into patches according to the latitude, and up-scales these patches separately. However, the information connection is blocked between adjacent patches. SphereSR~\cite{yoon2022spheresr} introduces a spherical local implicit image function (SLIIF) and a novel feature extraction module to leverage the information from arbitrary projection types, but introduced extremely high computational complexity. Recently, OSRT~\cite{yu2023osrt} introduced a distortion-aware transformer to address dimension-related distortions in ERP images, which achieves the state-of-the-art performance. Besides, fisheye downsampling and pseudo-ERP image generation method are proposed in OSRT to simulate real-world settings and alleviate network overfitting. However, the position information is still unclear in OSRT, and only spatial information is considered in the whole pipeline, which constraints the final performance.
Consequently, several attempts have been made to address 360° SR problems. LAU-Net~\cite{deng2021lau} divides the entire ERP image into patches based on latitude and upscales them separately. However, this approach obstructs the information connection between adjacent patches. SphereSR~\cite{yoon2022spheresr} introduces a Spherical Local Implicit Image Function (SLIIF) and a novel feature extraction module to leverage information from arbitrary projection types, but this results in extremely high computational complexity. Recently, OSRT~\cite{yu2023osrt} presented a distortion-aware transformer to address dimension-related distortions in ERP images, achieving state-of-the-art performance. Additionally, OSRT proposes fisheye downsampling and pseudo-ERP image generation methods to simulate real-world settings and mitigate network overfitting. However, position information remains unclear in OSRT, and only spatial information is considered throughout the pipeline, which constrains the final performance.

% To address the above problems, we propose a two-stage framework, combing two super-resolution networks and a same-resolution enhancement network. Specifically, two models are employed in the first stage (named model A and model B, separately). Model A is designed based on the architecture of HAT, incorporating proposed omnidirectional position-awared deformable blocks (OPDB) and fourier upsampling, while model B adds a spatial frequency fusion (SFF) module to model A. In the second stage, we perform a same-resolution enhancement based on the structure of model A with a pixel unshuffle operation at the beginning of the network. Besides, many strategies of data augmentation and training are conducted to prove the final restoration performance. In summary, our contributions include:
To address the aforementioned issues, we propose a two-stage framework that combines two super-resolution networks and a same-resolution enhancement network. Specifically, two models are employed in the first stage (referred to as model A and model B, respectively). Model A is designed based on the HAT architecture, incorporating proposed omnidirectional position-aware deformable blocks (OPDB) and Fourier upsampling, while model B adds a spatial frequency fusion (SFF) module to model A. In the second stage, we perform a same-resolution enhancement based on model A's structure, incorporating a pixel unshuffle operation at the beginning of the network. Moreover, various data augmentation and training strategies are implemented to improve the final restoration performance. In summary, our contributions include:

\begin{figure*}[t!]
  \centering
  \includegraphics[width=0.8\textwidth]{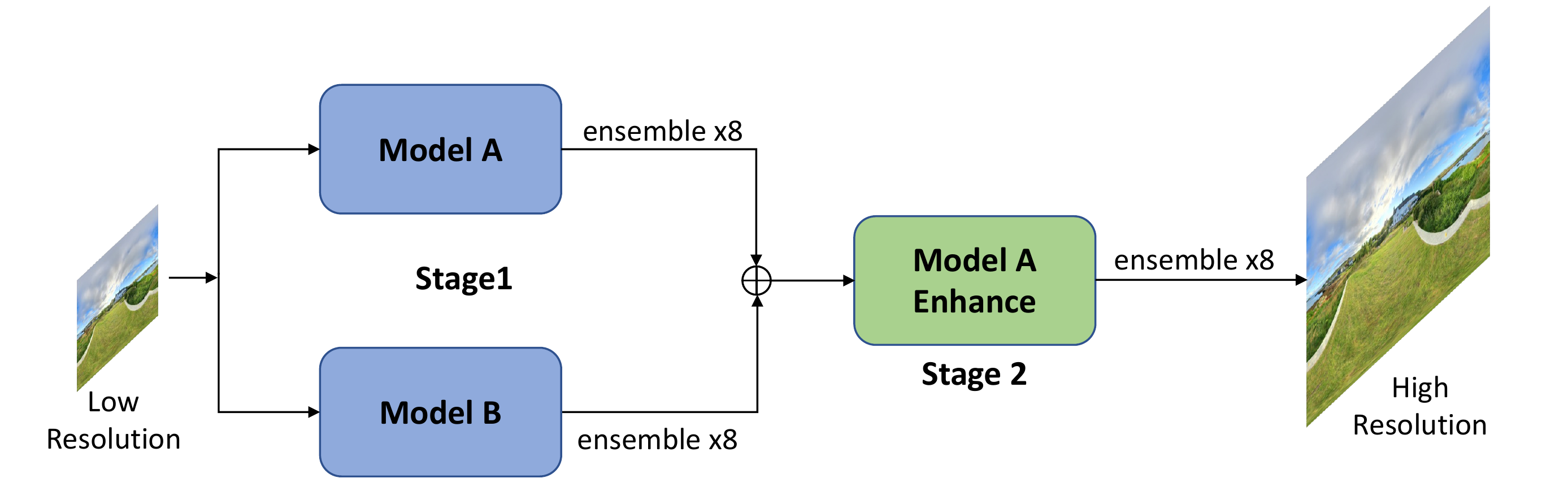}
  \caption{The overall pipeline of proposed two-stage framework, in which two $\times4$ SR models are employed in stage 1, while stage 2 performs a same-resolution enhancement.}
  \label{fig:overall_pipeline}
  \vspace{-0.2cm}
\end{figure*}

% \begin{itemize}
%     \item We propose a two-stage framework to obtain high-resolution images in the first stage and perform same-resolution enhancement in the second stage, which preserves more image details and eliminates artifacts.
%     \item We propose a novel omnidirectional position-awared deformable block (OPDB), which better leverages the position encoding information and ERP geometric properties.  
%     \item We introduce a spatial frequency fusion module (SFF) and Fourier upsampling to expore information in frequency domain.  
%     \item Our proposed method achieves superior performance, and wins the NTIRE 2023 challenge of 360° omnidirectional super resolution.
% \end{itemize}
\begin{itemize}
    \item We propose a two-stage framework that obtains high-resolution images in the first stage and performs same-resolution enhancement in the second stage, effectively preserving more image details and eliminating artifacts.
    \item We introduce a novel omnidirectional position-aware deformable block (OPDB), which efficiently leverages position encoding information and ERP geometric properties.
    \item We implement a spatial frequency fusion module (SFF) and Fourier upsampling to explore information in the frequency domain.
    \item Our proposed method achieves superior performance and wins the NTIRE 2023 challenge of 360° omnidirectional super-resolution~\cite{cao2023ntire}.
\end{itemize}

%------------------------------------------------------------------------
\section{Related Work}
\label{sec:related work}

%-------------------------------------------------------------------------
\subsection{Single Image Super-Resolution (SISR)}

Since the introduction of deep learning to single-image super-resolution tasks by SRCNN~\cite{dong2015image}, it has outperformed many traditional algorithms. Consequently, various CNN architectures have been extensively investigated by researchers to further improve the performance of image super-resolution algorithms. For example, EDSR~\cite{lim2017enhanced} initially employed residual blocks without batch normalization as the fundamental building blocks, forming a deeper super-resolution network. RDN~\cite{zhang2018residual} combined residual blocks with dense connections, introducing residual dense blocks. RCAN~\cite{zhang2018image} integrated channel attention into residual blocks, proposing residual attention modules and deepening the network. %In order to achieve superior perceptual quality in super-resolution reconstructions, GANs have also been introduced to the realm of image super-resolution.

Recently, Vision Transformers have overcome the inductive biases inherent in CNNs and effectively modeled long-range dependencies, achieving optimal performance in numerous high-level visual tasks. ViT-like structures have also been applied to low-level tasks. For instance, IPT~\cite{chen2021pre} proposed a network structure akin to ViT, pre-trained on large-scale datasets and multiple different low-level tasks, yielding impressive performance. SwinIR~\cite{liang2021swinir} adopted the Swin Transformer's structure, using a shifted window mechanism to model long-range dependencies and achieving enhanced performance with fewer parameters. EDT~\cite{li2021efficient} further improved single-image super-resolution performance by employing self-attention mechanisms and multi-related-task pre-training strategies.

HAT\cite{chen2205activating} combines self-attention, channel attention, and a novel overlapping cross-attention mechanism, introducing the residual hybrid attention groups (RHAG), which are composed of hybrid attention blocks (HAB) and overlapping cross-attention blocks (OCAB). This approach activates more pixels to facilitate reconstruction. In contrast to previous methods, HAT employs same-task pre-training on large-scale datasets, demonstrating the effectiveness of this strategy. Additionally, HAT expands the model's scale, establishing new state-of-the-art benchmarks for the single-image super-resolution task.

\begin{figure*}[h!]
  \centering
  \includegraphics[width=0.9\textwidth]{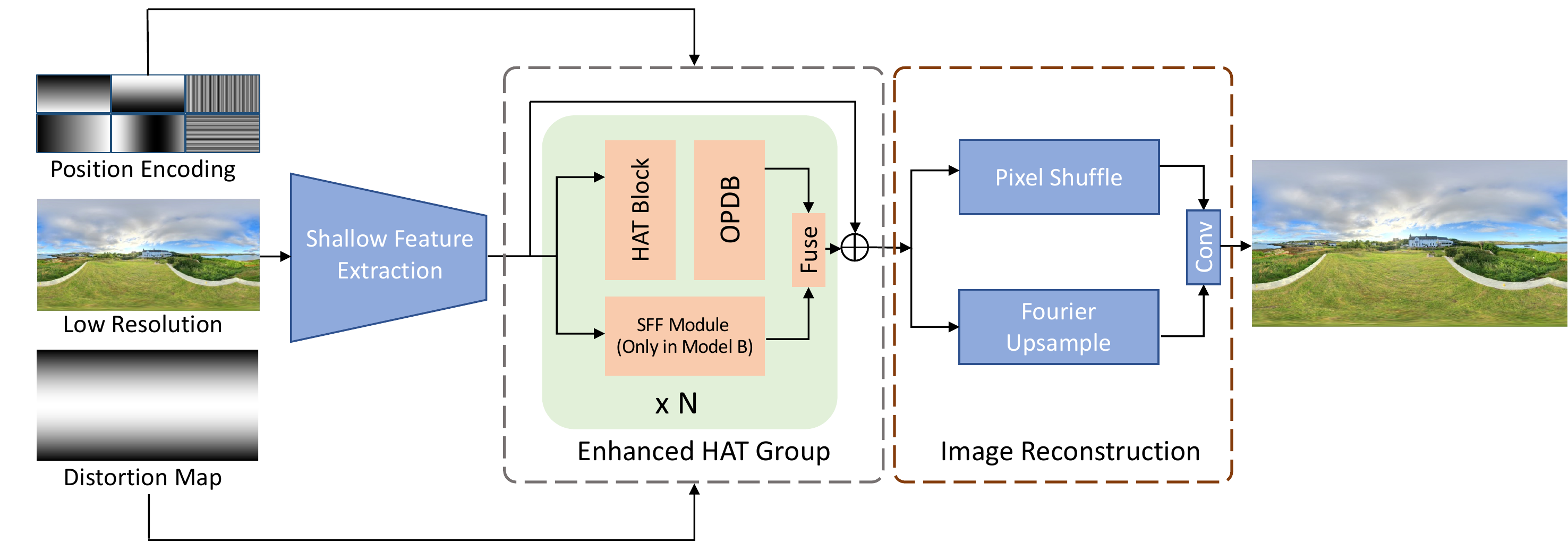}
  \caption{The network architecture of our proposed model A.}
  \label{fig:net_arch}
\end{figure*}

%-------------------------------------------------------------------------
\subsection{Omnidirectional Image Super-Resolution (ODISR)}

Initial research on omnidirectional image super-resolution (ODISR) concentrated on stitching and optimizing multiple low-resolution omnidirectional images using various projection types, such as spherical and hyperbolic. To assess the efficacy of omnidirectional image super-resolution qualitatively, Sun et al.\cite{sun2017weighted} proposed a weighted-to-spherically-uniform quality evaluation method (WS-PSNR) for spheres. Recently, GANs have been incorporated into ODISR, with models operating on planar images, fine-tuning existing SISR models using L1 loss\cite{fakour2018360} or GAN loss~\cite{zhang2020toward}, and employing WS-SSIM to evaluate model performance. LAU-Net~\cite{deng2021lau} identified pixel density non-uniformity in ERP omnidirectional images, prompting numerous studies to develop dedicated base networks addressing this issue. LAU-Net~\cite{deng2021lau} partitions the entire ERP image into latitude-related patches manually, learning ERP distortions within distinct latitude ranges. Instead of processing the whole ERP image end-to-end, LAU-Net separately processes non-overlapping patches of varying latitudes, resulting in discontinuities throughout the ERP image. Nishiyama et al. incorporated the area stretching ratio as an additional condition input to the network; however, this necessitates modifications to existing SISR backbone networks. SphereSR introduces an algorithm capable of handling omnidirectional image super-resolution for arbitrary projection types. Specifically, SphereSR\cite{yoon2022spheresr} proposes a feature extraction module that extracts features on a spherical surface from various projection types (such as CP, ERP, and polyhedra). Based on the extracted spherical features, SphereSR employs a spherical local implicit image function (SLIIF) to predict RGB values corresponding to spherical coordinates, yielding high-resolution reconstruction results for arbitrary projection types. Nonetheless, ERP remains the most commonly utilized projection type for omnidirectional image editing, transmission, and display.

In realistic scenarios, omnidirectional images (ODIs) are typically captured using two or more fisheye lenses, leading to distortions during the fisheye projection process. Recognizing this, Yu et al.\cite{yu2023osrt} proposed a degradation process—Fisheye downsampling—that emulates the imaging process in real-world settings. To more effectively address dimension-related distortions in equirectangular projection (ERP) images, Yu et al. introduced a Distortion-aware Transformer designed to adaptively perform super-resolution on omnidirectional images. Notably, panoramic image datasets tend to be smaller in size compared to conventional 2D images. To mitigate network overfitting, OSRT synthesizes pseudo-panoramic images from 2D images during training.

In this context, the OSRT approach aims to address the challenges inherent in omnidirectional image super-resolution, which differ from those in traditional 2D image super-resolution. By simulating real-world distortions during the degradation process and incorporating a Distortion-aware Transformer, this method is more adept at managing the distinctive features of omnidirectional images. Furthermore, employing synthesized pseudo-panoramic images alleviates the problem of limited panoramic image datasets, mitigating overfitting and enhancing the model's performance.

\begin{figure}[t!]
    \centering
    \setlength{\tabcolsep}{0.6pt}
    \begin{tabular}{cc}
         (a) Our simulated LR & (b) GT of LR \\
         \includegraphics[width=0.48\linewidth]{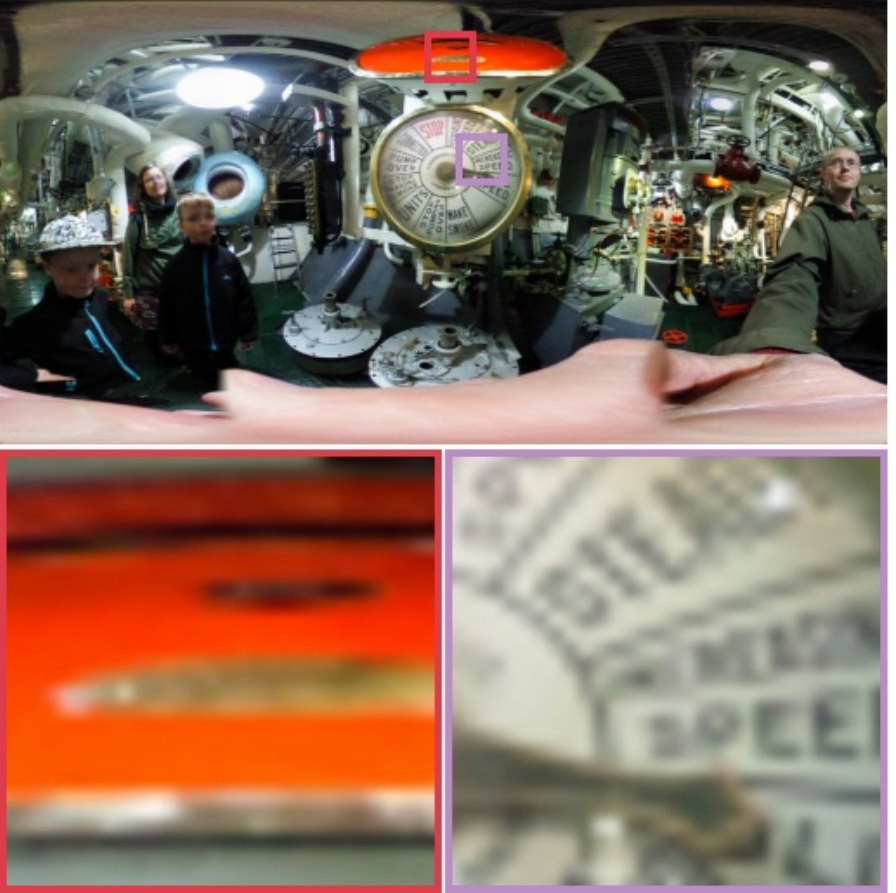} &
         \includegraphics[width=0.48\linewidth]{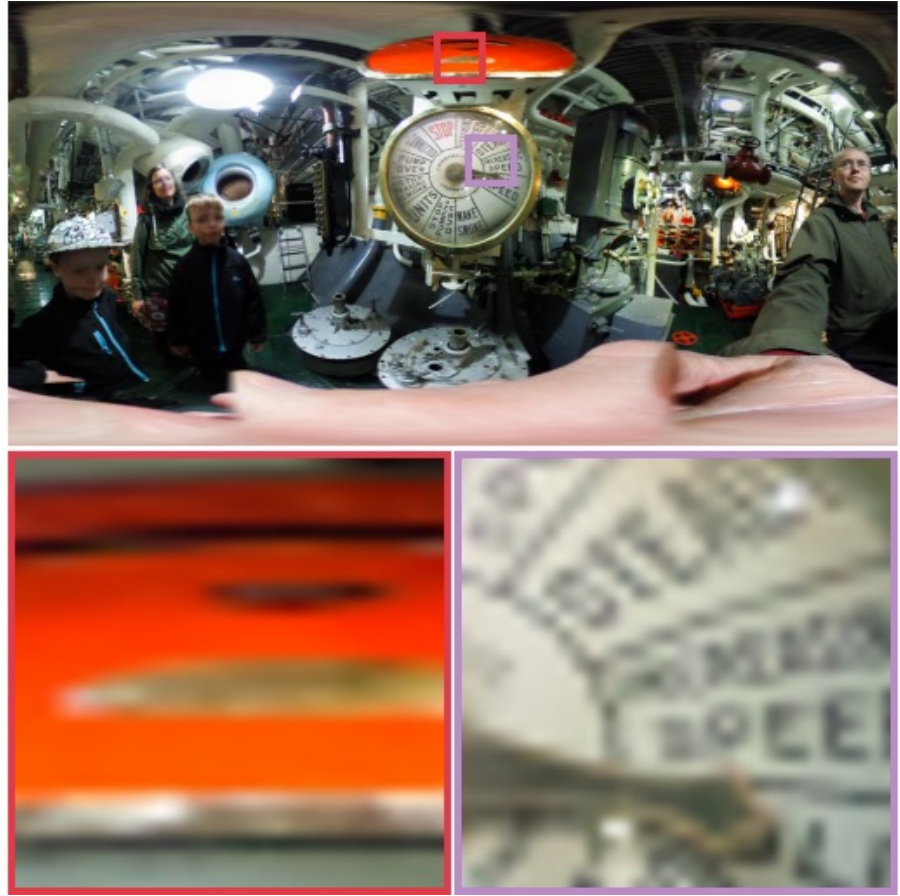}
    \end{tabular}
    \vspace{-5pt}
    \caption{Visualization comparisons of (a) our simulated LR images with applying our degradation network to one HR image in Flickr360 and (b) the corresponding groudtruth LR image.}
    \label{fig:lbo_lr}
    \vspace{-15pt}
\end{figure}

%------------------------------------------------------------------------
\section{Method}

\subsection{Simulating Data Degradation}
% Transformer-based models require extensive data input for training to achieve improved performance~\cite{chen2020pre}, we collected numerous panoramic video data from YouTube to train our model. Since the competition organizers did not provide HR-LR degradation settings, we drew inspiration from AnimeSR's LBO~\cite{animeSR} module and employed a degradation network identical to LBO to learn the mapping from HR to LR on the training set. The converged degradation model achieved a WS-PSNR~\cite{sun2017weighted} of 43 when tested on the Flickr360 validation set. Subjectively, as shown in Fig.~\ref{fig:lbo_lr}, the generated pseudo-LR images exhibit higher clarity in the polar regions compared to LR, but lower clarity in regions with low dimensions.
% \subsection{Proposed Two-stage Framework}
Transformer-based models necessitate substantial data input for training to enhance performance~\cite{chen2021pre}. Therefore, we amassed a vast collection of panoramic video data from YouTube\footnote{\url{www.youtube.com}} to train our model. Due to the lack of HR-LR degradation settings provided by the competition organizers, we adopted the LBO module from AnimeSR~\cite{animeSR} and implemented a degradation network identical to LBO to learn the mapping from HR to LR on the training set. The fully converged degradation model achieved a WS-PSNR~\cite{sun2017weighted} of 43 when evaluated on the Flickr360 validation set. As illustrated in Fig.~\ref{fig:lbo_lr}, the generated pseudo-LR images subjectively exhibit increased clarity in polar regions compared to LR, yet diminished clarity in regions with low dimensions.
\subsection{Proposed Two-Stage Framework}

% 我们提出一个二阶段SR模型来做360imageSR，第一阶段是SRx4模型，第二阶段是同分辨率增强模型。在第一阶段，我们采用了两个模型(model A and model B)进行模型投票，model A是基于HAT的模型结构加入了OPDB和傅立叶上采样，model B在model A上加入了频域模块，事实上，只有model A就可以让我们赢得这次比赛的冠军。在二阶段，采用基于model A的结构和权重参数，在前置增加pixel unshuffle对输入图像进行4x下采样，这样可以得到与输入相同分辨率的结果。

% We propose a two-stage SR model for 360-image SR, as Fig.~\ref{fig:overall_pipeline} shows, with the first stage consisting of an SRx4 model and the second stage comprising a same-resolution enhancement model. In the first stage, we employ two models (model A and model B) for model ensemble. Model A is based on the HAT model structure, incorporating OPDB and Fourier upsampling, while model B adds a spatial frequency fusion module(SFF) to model A. In fact, model A alone would suffice to secure our victory in this competition. In the second stage, we utilize a structure based on model A and its weight parameters, adding a pixel unshuffle at the beginning to downsample the input image by 4x, thereby yielding a result with the same resolution as the input.
We propose a two-stage super-resolution (SR) model for 360-degree image SR, as illustrated in Fig.~\ref{fig:overall_pipeline}. The first stage comprises an SRx4 model, while the second stage features a same-resolution enhancement model. In the first stage, we utilize two models (Model A and Model B) for ensemble purposes. Model A is built on the Hybrid Attention Transformer (HAT) architecture, integrating Orthogonal Projection Depthwise Block (OPDB) and Fourier upsampling, whereas Model B incorporates a Spatial Frequency Fusion (SFF) module into Model A. In fact, Model A alone is sufficient to ensure our victory in this competition. In the second stage, we employ a structure based on Model A and its weight parameters, introducing a pixel unshuffle at the beginning to downsample the input image by a factor of four, ultimately producing a result with the same resolution as the input.

\subsection{First Stage with OPDB}

\begin{figure}[t]
  \centering
  \includegraphics[width=0.45\textwidth]{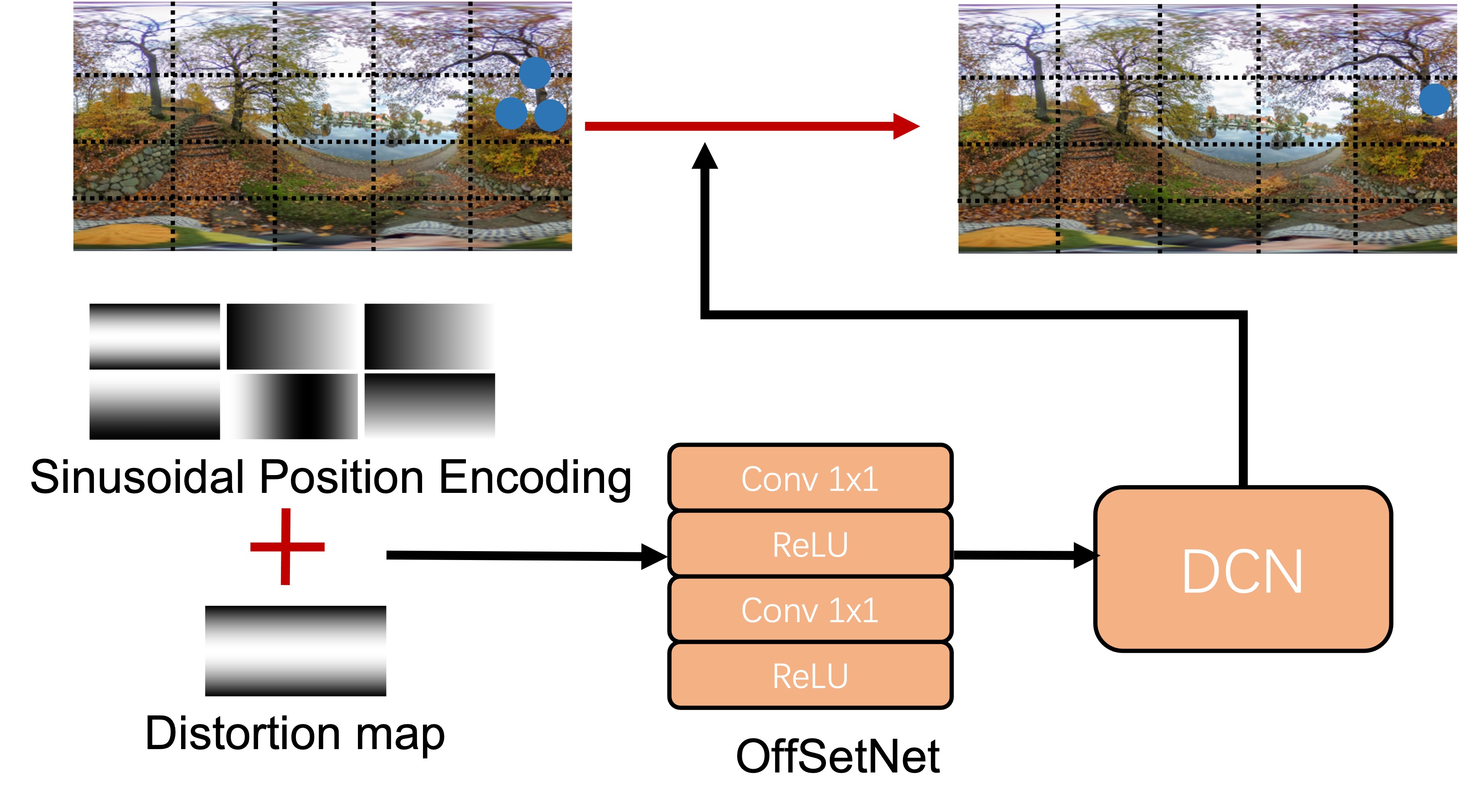}
  \caption{OPDB Module.}
  \label{fig:opdb_arch}
  \vspace{-7pt}
\end{figure}
% In stage I, we based our network structure on HAT and added OPDB after each RHAG, along with a frequency domain module for spatial frequency fusion in Fig.~\ref{fig:net_arch}. Finally, Fourier upsampling was added to the upsampling module~\cite{DFUS}. Specifically, inspired by DACB in~\cite{yu2023osrt} and BUSIFusion~\cite{busifusion}, a novel Omnidirectional Position-aware  Deformable Block (OPDB) was proposed, which combines dimensional information and position encoding information for 360-degree images. As shown Fig.~\ref{fig:opdb_arch}, OPDB uses Sinusoidal Position Encoding~\cite{vaswani2017attention} to encode the position of ERP projections: absolute positions are represented using sine and cosine functions, and relative positions are obtained by multiplying the two. This design allows positional encoding to be linearly represented by position, reflecting its relative position relationship.
Model A and Model B were combined through model ensemble to form the first stage. In Model A, we based our network structure on HAT and added OPDB after each RHAG, along with a frequency domain module for spatial frequency fusion in Fig.~\ref{fig:net_arch}. Finally, Fourier upsampling was added to the upsampling module~\cite{DFUS}. Specifically, inspired by DACB in~\cite{yu2023osrt} and BUSIFusion~\cite{busifusion}, a novel Omnidirectional Position-aware  Deformable Block (OPDB) was proposed, which combines dimensional information and position encoding information for 360-degree images. As shown Fig.~\ref{fig:opdb_arch}, OPDB uses Sinusoidal Position Encoding~\cite{vaswani2017attention} and distortion map~\cite{yu2023osrt} to encode the position of ERP projections: absolute positions are represented using sine and cosine functions, and relative positions are obtained by multiplying the two. This design allows positional encoding to be linearly represented by position, reflecting its relative position relationship:

\begin{equation}
PE_{(pos,2i)}=sin(\frac{pos}{10000^{2i/d_{model}}}),
\end{equation}

\begin{equation}
PE_{(pos,2i+1)}=cos(\frac{pos}{10000^{2i/d_{model}}}).
\end{equation}
Here, $pos$ denotes the position in the input sequence, which is the coordinate on the ERP image. We encode the latitude and longitude of the ERP image separately using the above encoding method. $i$ represents the index of the sine function, and $d_{model}$ signifies the dimension of the model. In this manner, each position is encoded into a $d_{model}$-dimensional vector, with each dimension corresponding to a sine function. When encoding the position of Equirectangular Projection (ERP) representations, the ERP representation can be regarded as a sequence, and Sinusoidal Position Encoding can be employed to encode its position. Consequently, each ERP representation is encoded into a $d_{model}$-dimensional vector, which encompasses its position information in the sequence. This encoding technique aids the model in better understanding the relationships among different ERP representations, thereby enhancing the model's performance. The positional encoding information is transmitted to the offsetNet, allowing the deformable convolution to utilize this offset information to adjust the convolution kernel based on the spherical position correlation. In this manner, the spherical coordinate information can be aggregated through deformable convolution, achieving a larger receptive field and superior reconstruction effect.

\begin{figure}[t!]
  \centering
  \includegraphics[width=0.45\textwidth]{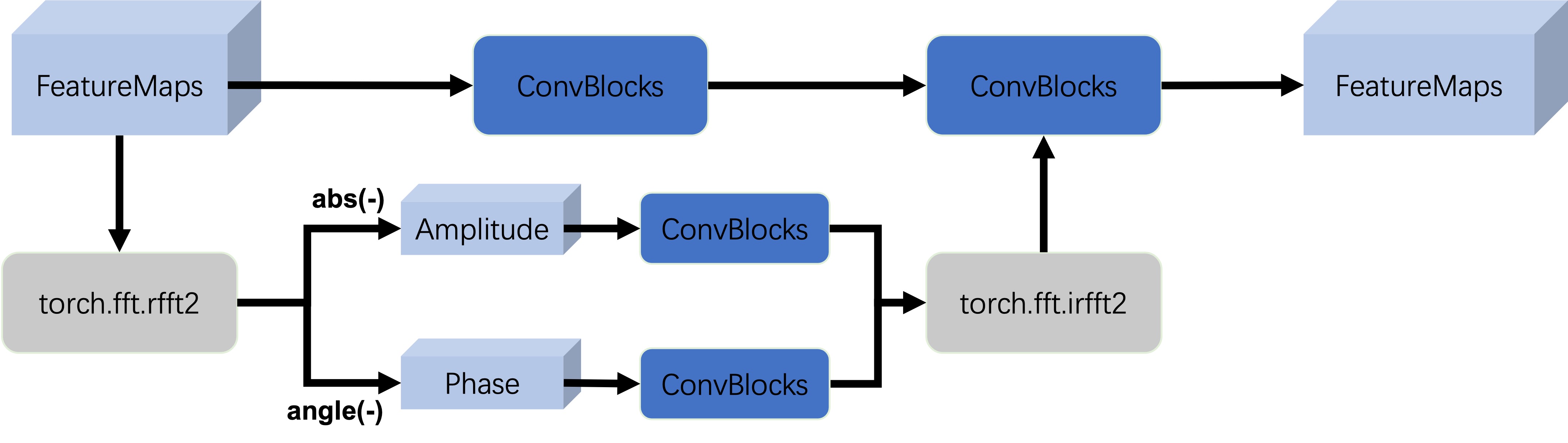}
  \caption{SFF Module.}
  \label{fig:sff_arch}
  \vspace{-7pt}
\end{figure}

% The above modules constitute Model A, and the following is the SSF module newly added to Model B.
% The spatial frequency fusion module is shown in the Fig. \ref{fig:sff_arch}, which performs Fourier transform on the feature map, extracts features from the real and imaginary parts separately, and finally performs ifft transform to fuse the feature map with spatial domain information.
The aforementioned modules constitute Model A, while the subsequent Spatial Frequency Fusion (SSF) module is a novel addition to Model B. Inspired by the work of Zhou et al.~\cite{zhou2022spatial}, the SSF module, as depicted in Fig.~\ref{fig:sff_arch}, conducts a Fourier transform on the feature map, extracts features from the amplitude and phase components individually. After that, the processed amplitude and phase components are transformed back into the complex domain. Finally, an inverse Fast Fourier Transform (IFFT) is performed on the complex feature to merge the feature map with spatial domain information.

\begin{figure}[t!]
	%\newlength-4mm
	%\setlength{-4mm}{-0.4cm}
	\vspace{-6mm}
	\scriptsize
	\centering
	\begin{tabular}{l}
		\hspace{-0.42cm}
		\begin{adjustbox}{valign=t}
			\begin{tabular}{c}
				\includegraphics[width=0.46\columnwidth]{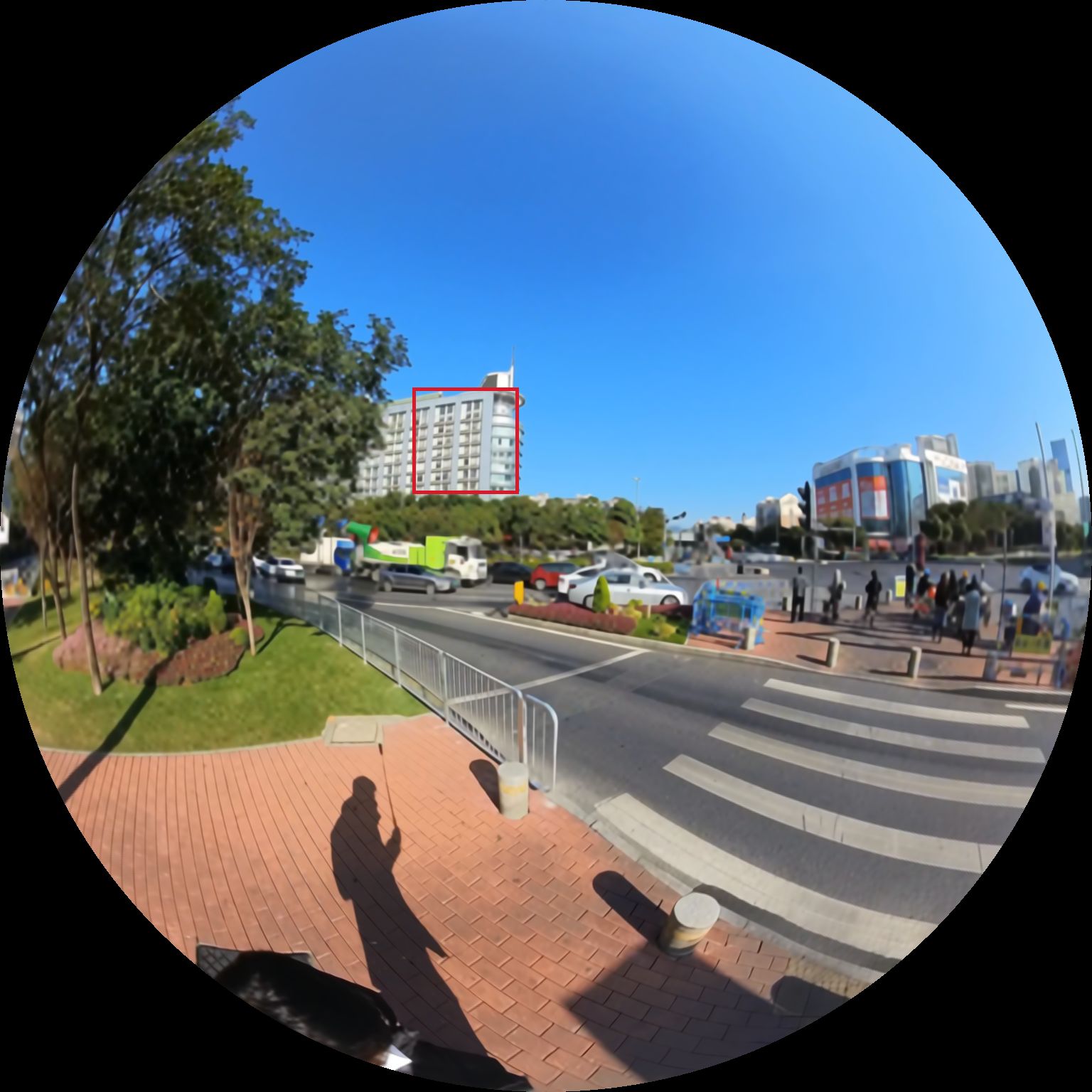}
				\\
				Flickr360 ($\times$4): 113
				\\
				Fisheye (Vertical, Right)
			\end{tabular}
		\end{adjustbox}
		\hspace{-3mm}
		\begin{adjustbox}{valign=t}
			\begin{tabular}{cc}
				\includegraphics[width=0.23\columnwidth]{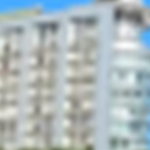} \hspace{-3mm} &
				\includegraphics[width=0.23\columnwidth]{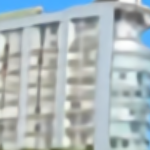} \hspace{-1mm} 
				\\
				Bicubic  \hspace{-3mm} &
				EDSR \cite{lim2017enhanced} \hspace{-1mm} 
				\\
				\includegraphics[width=0.23\columnwidth]{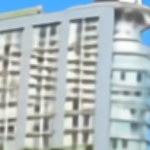} \hspace{-3mm} &
				\includegraphics[width=0.23\columnwidth]{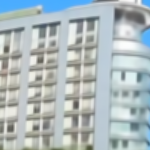} \hspace{-1mm} 
				
				\\ 
				OSRT\cite{yu2023osrt}  \hspace{-3mm} &
				OPDN (Ours) \hspace{-1mm} 
				
			\end{tabular}
		\end{adjustbox}
		\vspace{1mm}
		
		\\ 
		\hspace{-0.42cm}
		\begin{adjustbox}{valign=t}
			\begin{tabular}{c}
				\includegraphics[width=0.46\columnwidth]{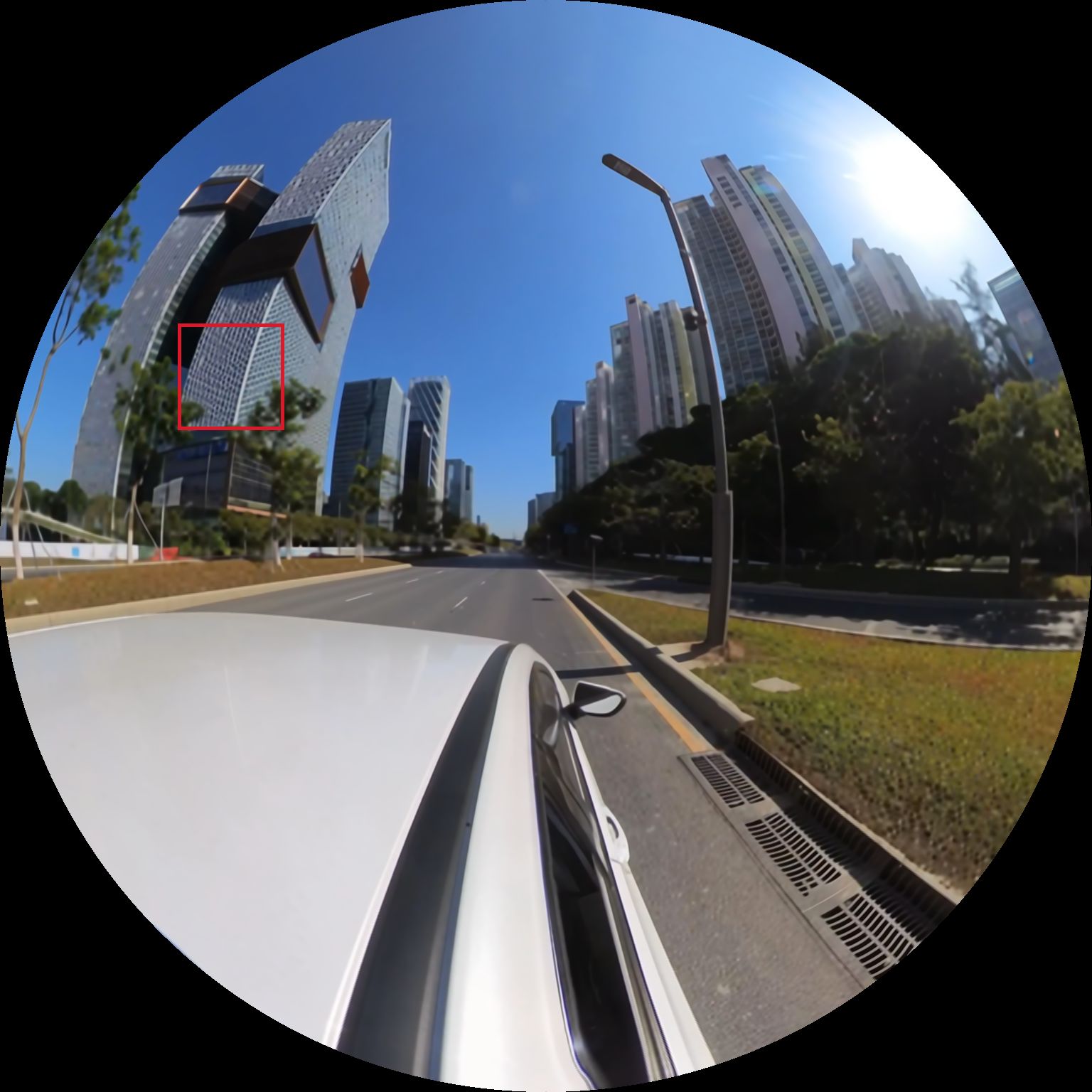}
				\\
				Flickr360 ($\times$4): 120
				\\
				Fisheye (Vertical, Right)
			\end{tabular}
		\end{adjustbox}
		\hspace{-3mm}
		\begin{adjustbox}{valign=t}
			\begin{tabular}{cc}
				\includegraphics[width=0.23\columnwidth]{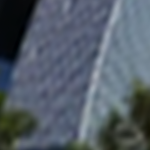} \hspace{-3mm} &
				\includegraphics[width=0.23\columnwidth]{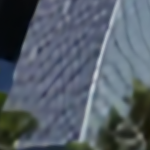} \hspace{-1mm} 
				\\
				Bicubic  \hspace{-3mm} &
				EDSR \cite{lim2017enhanced} \hspace{-1mm} 
				\\
				\includegraphics[width=0.23\columnwidth]{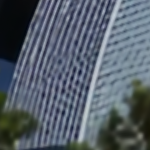} \hspace{-3mm} &
				\includegraphics[width=0.23\columnwidth]{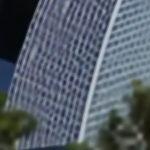} \hspace{-1mm} 
				
				\\ 
				OSRT\cite{yu2023osrt} \hspace{-3mm} &
				OPDN (Ours)\hspace{-1mm} 
				
			\end{tabular}
		\end{adjustbox}
		
	\end{tabular}
	\vspace{-3mm}
	\caption{Visual comparisons for SR of Fisheye images.}
	\label{fig:fisheye}
    \vspace{-17pt}
\end{figure}

\subsection{Second Stage and Overall Strategy}

% For the second stage model, we added a pixel unshuffle operation before the model to maintain the same resolution for input and output, based on the weights of Model A. The training data for the second stage is generated by inference on the training data using Model A, without self-ensemble. We found that using self-ensemble during the second stage training led to performance loss, as shown in the table.
In the second stage model, we introduced a pixel unshuffle operation before the model, based on the weights of Model A, to ensure that the input and output resolutions remain consistent. The training data for the second stage is generated by performing inference on the training data using Model A, without employing self-ensemble. We discovered that utilizing self-ensemble during the second stage training resulted in performance degradation, as demonstrated in the table.

% The overall training process is as follows:
% Stage I: First, we fine-tune the pre-trained HAT-L model on the competition data, then fine-tune it on the generated degraded 360-degree data, and finally fine-tune it again on the competition data, with the window size in HAB increased from 16 to 32. Two branches are derived from this process: model A without the SSF module, and model B with it, which are used for model ensemble.
The overall training process comprises the following steps:

Stage I: Initially, we fine-tune the pre-trained HAT-L model on the competition data, followed by fine-tuning it on the generated degraded 360-degree data, and finally fine-tuning it once more on the competition data, while increasing the window size in the Hybrid Attention Block (HAB) from 16 to 32. This process yields two branches: Model A, which does not include the Spatial Frequency Fusion (SSF) module, and Model B, which incorporates the SSF module. Both models are employed for ensemble purposes.

% Stage II: We follow the approach described in Section 3.4.

% Testing stage: We first perform inference on Model A and Model B from stage I and use a new self-ensemble x8 technique. Considering the characteristics of ERP images, we only perform horizontal and vertical flipping, roll 1/4 of the width, and then perform horizontal and vertical flipping again. We then average the results of Model A and Model B, and obtain the final result from inference using the second stage model.
Stage II: We adhere to the approach delineated in Section 3.4.

\begin{figure}[h!]
  \centering
  \includegraphics[width=0.45\textwidth]{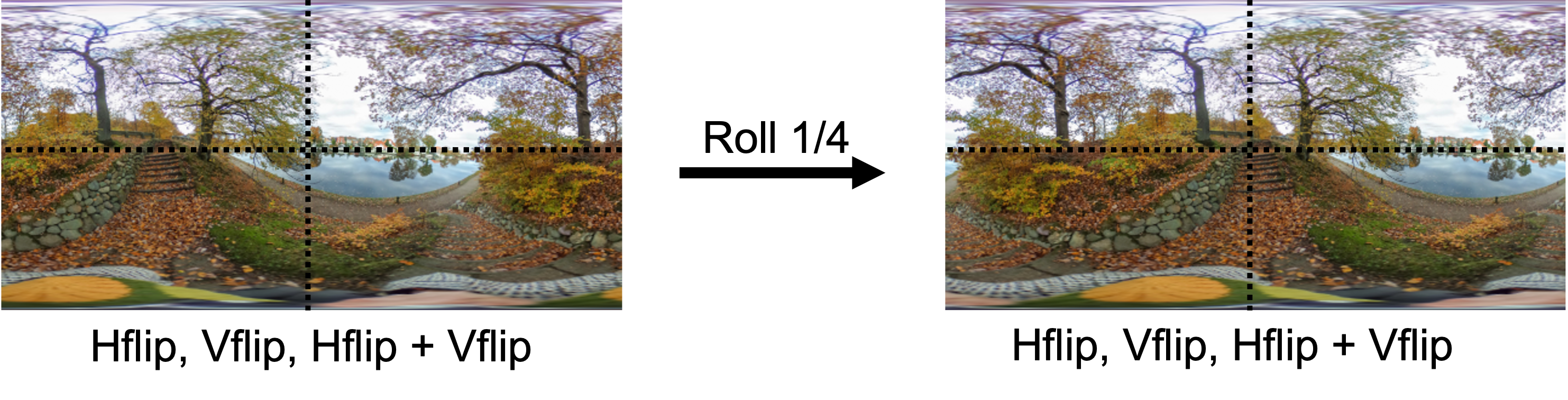}
  \vspace{-8pt}
  \caption{Self-ensemble strategy for ERP images.}
  \label{fig:self_ensemble}
\end{figure}

Testing stage: Initially, we conduct inference on Models A and B from Stage I and apply a novel self-ensemble x8 technique. Taking into account the characteristics of Equirectangular Projection (ERP) images, in Fig~\ref{fig:self_ensemble}, we perform horizontal and vertical flipping, roll 1/4 of the width (We have tested other parameters, and only 1/4 showed improvement.), and subsequently conduct horizontal and vertical flipping once more. Next, we average the results of Models A and B, and derive the final result from inference using the second stage model.

\vspace{-5pt}
\vspace{-20pt}
\section{Experiments}

\subsection{Datasets}

We utilized two datasets in our experiments. The first dataset is the Flickr360 dataset, which serves as the official data for the NTIRE 2023 competition and is used for model training and testing. In addition to Flickr360, we collected an extra 260 high-resolution panoramic videos from YouTube\footnote{\url{https://github.com/sxpro/OPDN/blob/main/youtube_video_data_list.txt}}. We extracted all the I-frames from these videos, converted frames in non-ERP projection formats (such as EAC, Cubemap, and Stereo EAC) to the ERP projection format, and subsequently downsampled these I-frames to a resolution of $2048\times1024$. We manually removed low-quality frames or those influenced by transition effects in the videos. Lastly, we employed the LBO degradation network to downsample the selected approximately 7,000 frames, obtaining their corresponding low-resolution images.

\subsection{Implementation Detail}
% For stage I, we first fine-tune the official pre-trained HAT-L model for 450K iterations with Charbonnier loss. Adam optimizer is adopted with an initial learning rate of $1 \times 10^{-4}$. We adopt the MultiStepLR strategy for learning rate adjustment, gradually reducing it to $1 \times 10^{-6}$. The same training strategy is used for all training in stage I. Finally, we fine-tune our model with L2 loss for 10K iterations.

% For stage II, we train our model with L2 loss for 300K iterations, using an initial learning rate of $5 \times 10^{-5}$.
% All experiments are conducted with four NVIDIA A100 GPUs.
For Stage I, we initially fine-tune the official pre-trained HAT-L model for 450K iterations using the Charbonnier loss. The Adam optimizer is employed with an initial learning rate of $1 \times 10^{-4}$. We implement the MultiStepLR strategy for learning rate adjustment, progressively reducing it to $1 \times 10^{-6}$. The same training strategy is applied to all training instances in Stage I. Ultimately, we fine-tune our model using the L2 loss for 10K iterations.

For Stage II, we train our model using the L2 loss for 300K iterations, with an initial learning rate of $5 \times 10^{-5}$. All experiments are conducted using four NVIDIA A100 GPUs.

\begin{table}
    \caption{Final results of the NTIRE2023 Challenge on 360° Image SR track~\cite{cao2023ntire}.}
    \label{tab:ntire_results}
    \normalsize
    \centering
    \setlength{\tabcolsep}{4pt}

    \begin{tabular}{ccc}
        \toprule[.1em]
        Methods & WS-PSNR (Val) & WS-PSNR (Test) \\
        \midrule[.1em]
        \textbf{1st (Ours)} & \textbf{30.43} & \textbf{28.64}\\
        2nd & 30.20 & 28.49\\
        3rd & 30.04 & 28.28\\
        4th & 30.03 & 28.13\\
        5th & 29.87 & 28.11\\
        6th & 30.00 & 28.10\\
        7th & 28.32 & 27.65\\
        \bottomrule[.1em]
    \end{tabular}
    \vspace{-10pt}
\end{table}

\begin{table}
    \caption{ Quantitative results of PSNR on Flickr360 dataset.}
    \label{tab:quan}
    \normalsize
    \centering
    \setlength{\tabcolsep}{4pt}

    \begin{tabular}{ccc}
        \toprule[.1em]
        Methods & WS-PSNR (Val) & WS-PSNR (Test) \\
        \midrule[.1em]
        Bicubic & 27.45 & 25.74\\
        EDSR-M~\cite{lim2017enhanced} & 29.18 & 27.30\\
        SwinIR~\cite{liang2021swinir} & 29.75 & 27.86\\
        \midrule[.1em]
        OSRT~\cite{yu2023osrt} & 30.05 & -\\
        OPDN (Ours) & \textbf{30.43} & \textbf{28.64}\\
        \bottomrule[.1em]
    \end{tabular}
    \vspace{-10pt}
\end{table}

\subsection{Quantitative Results}
% We compare our OPDN with previous SISR methods~\cite{li2021efficient, liang2021swinir} and ODISR method~\cite{yu2023osrt}. For fair comparison, we fine-tune OSRT~\cite{yu2023osrt} on the Flickr360 training set. Since the test set is not yet publicly available, the WS-PSNR of OSRT on test sets is not obtained. As shown in Table~\ref{tab:quan}, our OPDN outperforms pervious method by 0.38dB on WS-PSNR.
We compare our Omnidirectional Position-aware Deformable Network (OPDN) with previous Single Image Super-Resolution (SISR) methods~\cite{li2021efficient, liang2021swinir} and the Omnidirectional Deep Image Super-Resolution (ODISR) method~\cite{yu2023osrt}. For fair comparison, we fine-tune OSRT~\cite{yu2023osrt} on the Flickr360 training set. Since the test set is not yet publicly available, the WS-PSNR of OSRT on test sets is not obtained. As shown in Table~\ref{tab:quan}, our OPDN surpasses the previous method by 0.38dB in terms of WS-PSNR.

\subsection{Qualitative Results}
% We present our results on the official validation set of NTIRE 2023 challenge (\textit{i.e.} Flickr360) in Fig~\ref{fig:visual_val}. It can be observed that our proposed OPDN recovers more reliable texture details which are not captured by OSRT\cite{yu2023osrt}. Besides, one can see that our OPDN recovers with better lines and stripes, fewer visible artifacts and less blurry effect.
We present our results on the official validation set of the NTIRE 2023 challenge (i.e., Flickr360) in Fig.~\ref{fig:visual_val}. It is evident that our proposed Omnidirectional Position-aware Deformable Network (OPDN) restores more reliable texture details, which are not captured by OSRT\cite{yu2023osrt}. Furthermore, our OPDN exhibits superior recovery of lines and stripes, with fewer visible artifacts and a reduced blurring effect. From Fig.~\ref{fig:fisheye}, one can see that OPDN can preserve the original structure when being projected to other projection types.

\begin{figure}[t!]
    \centering
    \includegraphics[width=1\linewidth]{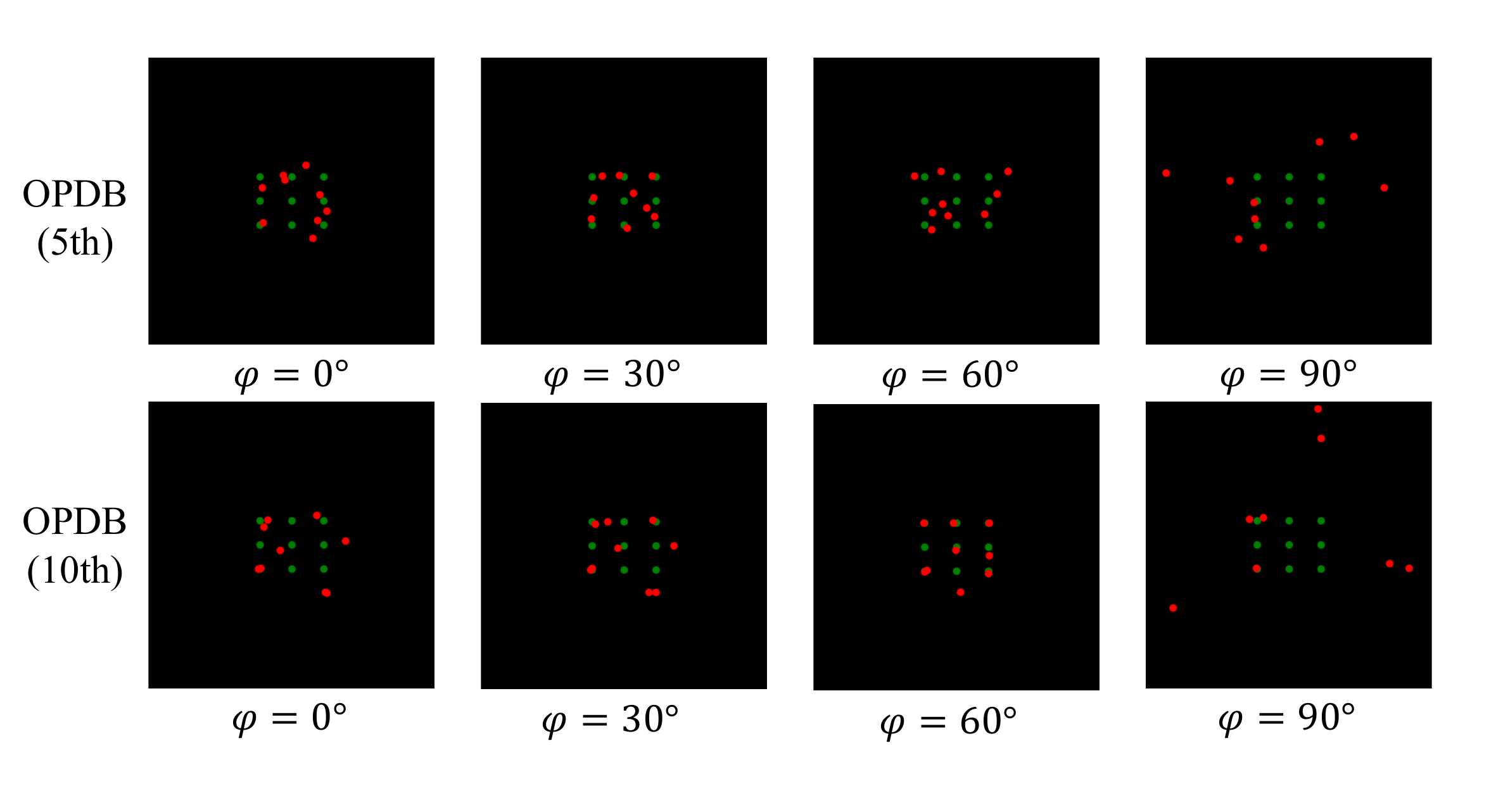}
     \vspace{-30pt}
    \caption{Visualizations of offset maps in OPDB. Reference and deformed points are depicted in \textcolor{green}{green} and \textcolor{red}{red}, respectively.}
    \label{fig:offset}
    \vspace{-7pt}
\end{figure}

\begin{figure}[t!]
    \centering
    \setlength{\tabcolsep}{0.6pt}
    \small
    % \begin{subfigure}[b]{0.3\textwidth}
    \subfloat[Flickr360 No.0307]{
    % \begin{minipage}[b]{0.4\textwidth}
        \includegraphics[width=0.16\textwidth,height=0.17\textheight]{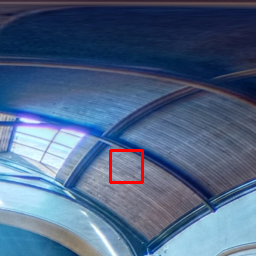}
        % \caption{LR Image}
    }
    \hfill
    % \end{minipage}
    % \end{subfigure}
    \subfloat[OPDN]{
    % \begin{minipage}[b]{0.4\textwidth}
        \begin{tabular}{ccccc}
             % OPDN w/o OPDB &
             \subfloat[OPDN w/o sinusoidal position encoding]{
             \includegraphics[width=0.6\linewidth]{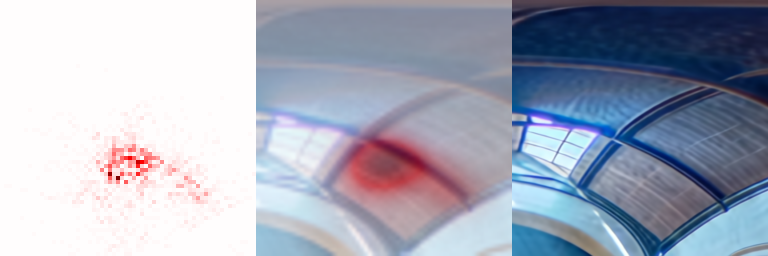}
             % \caption{OPDN w/o OPDB}
             }
             \\
             % OPDB &
             \includegraphics[width=0.6\linewidth]{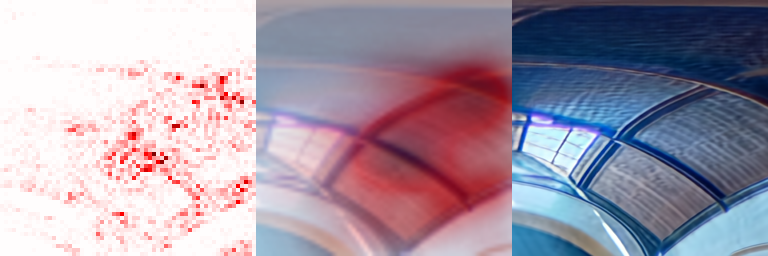}\\
        \end{tabular}
    % \end{minipage}
    }
    \vspace{-7pt}
    \caption{Results of LAM~\cite{gu2021interpreting} visualization. From left to right, (b) and (c) show the LAM contribution, area of contribution and SR results. The LAM maps illustrate the significance of each pixel in the input LR image with respect to the SR of the patch indicated by a red box. The LAM results demonstrate that OPDB can effectively adapt to changes in dimensions of 360-degree images and utilize a broader range of information to reconstruct SR results.}
    \label{fig:lam_vis}
    \vspace{-7pt}
\end{figure}

\subsection{Ablation Study}

\begin{table}[t]
  \caption{Ablation results in Flickr360 validation.}
  \vspace{-5pt}
  \label{tab:ablation}
  \centering
  \setlength{\tabcolsep}{4pt}
  \begin{tabular}{ccccc}
    \toprule[.1em]
    Method & WS-PSNR  \\
    \midrule[.1em]
    HAT-L imageNet pretrained & 30.15 \\
    + Fourier upsampling & 30.16\\
    + YouTube data & 30.17 \\
    + windowsize 32 & 30.17 \\
    + SSF & 30.18 \\
    + OPDB & 30.28 \\
    \bottomrule[.05em]
    OPDN & \textbf{30.37} \\
    \bottomrule[.1em]
  \end{tabular}
  \vspace{-10pt}
\end{table}

% The ablation experiments mainly compare the effectiveness of OPDB, SSF, Fourier upsampling and youtube data. The ablation experiments were conducted by fine-tuning the HAT-L ImageNet pre-trained weights with the same settings, and testing the PSNR on the validation set. As shown in Table~\ref{tab:ablation}, our proposed methods all achieve significant performance improvements.
The ablation experiments primarily evaluate the effectiveness of Omnidirectional Position-aware Deformable Block (OPDB), Spatial Spectral Fusion (SSF), Fourier upsampling, and YouTube data. These experiments are conducted by fine-tuning the HAT-L ImageNet pre-trained weights using identical settings and assessing the PSNR on the validation set. As shown in Table~\ref{tab:ablation}, our methods consistently achieve substantial performance improvements. Figs.~\ref{fig:offset} and~\ref{fig:lam_vis} illustrate that OPDB is capable of adapting to dimensional changes in 360-degree images and leveraging a wider range of information to reconstruct SR results.

\begin{table}[t]
  \caption{Ensemble results in Flickr360 dataset.}
  \label{tab:ablation_ensemble}
  \centering
  \resizebox{0.5\textwidth}{!}{
  \begin{tabular}{cccc}
    \toprule[.1em]
    Method & WS-PSNR(Val)  & WS-PSNR(Test) & Infer Time\\
    \midrule[.1em]
    HAT-L & 30.15 & - & 4.4s/per\\
    Model A & 30.37 & - & 4.5s/per\\
    Model B & 30.37 & - & 5s/per\\
    Model A + se x8 & 30.41 & 28.61 & 35s/per\\
    Model B + se x8 & 30.41 & 28.61 & 37s/per\\
    Stage1 & 30.42 & 28.62 & 66s/per\\
    Stage2 & \textbf{30.43} & \textbf{28.64} & 73s/per \\
    \bottomrule[.1em]
  \end{tabular}
  }
  \vspace{-10pt}
\end{table}

% \section{NTIRE 2023 Challenge}
% We participate in track1 in the NTIRE 2023 360° Omnidirectional Super Resolution. Quantitative results are presented in Table x. In the competition, the new self-ensemble is used in all stages, with the model-ensemble. In fact, Model A alone would have been sufficient to win the championship, but we strived for excellence and improved it by an additional 0.03 dB as shown in Table~\ref{tab:ablation_ensemble}.
% \section{Conclusion}
% In this paper, we propose an Omnidirectional Position-aware Deformable Network for 360-degree image super-resolution. Specifically, we introduce a two-stage framework, OPDB, which includes a frequency block and Fourier upsampling to improve the final performance of image enhancement. Our method achieves a good trade-off between enhancement performance and model complexity, and won the championship in the 360° Omnidirectional Super-Resolution category of NTIRE 2023.
\section{NTIRE 2023 Challenge}
We participate in Track 1 of the NTIRE 2023 360° Omnidirectional Super-Resolution Challenge~\cite{cao2023ntire}. Quantitative results are presented in Table~\ref{tab:ntire_results}. During the competition, the new self-ensemble is utilized in all stages, along with the model ensemble. In fact, Model A alone would have sufficed to secure the championship; however, we pursued excellence and achieved an additional 0.03 dB improvement, as shown in Table~\ref{tab:ablation_ensemble}. The testing condition is to infer LR images of size $512\times256$ on A100.

\section{Conclusion}
In this paper, we propose an Omnidirectional Position-aware Deformable Network for 360-degree image super-resolution. Specifically, we introduce a two-stage framework and OPDB, which incorporates a frequency block and Fourier upsampling to enhance the final performance of image improvement. Our method strikes a favorable balance between enhancement performance and model complexity, ultimately winning the championship in the 360° Omnidirectional Super-Resolution category of NTIRE 2023.

%%%%%%%%% REFERENCES
{\small
\bibliographystyle{ieee_fullname}
\bibliography{egbib}

\begin{thebibliography}{10}\itemsep=-1pt

\bibitem{ai2022deep}
Hao Ai, Zidong Cao, Jinjing Zhu, Haotian Bai, Yucheng Chen, and Ling Wang.
\newblock Deep learning for omnidirectional vision: A survey and new
  perspectives.
\newblock {\em arXiv preprint arXiv:2205.10468}, 2022.

\bibitem{cao2023ntire}
Mingdeng Cao, Chong Mou, Fanghua Yu, Xintao Wang, Yinqiang Zheng, Jian Zhang,
  Chao Dong, Ying Shan, Gen Li, Radu Timofte, et~al.
\newblock {NTIRE} 2023 challenge on 360° omnidirectional image and video
  super-resolution: Datasets, methods and results.
\newblock In {\em Proceedings of the IEEE/CVF Conference on Computer Vision and
  Pattern Recognition Workshops}, 2023.

\bibitem{chen2021pre}
Hanting Chen, Yunhe Wang, Tianyu Guo, Chang Xu, Yiping Deng, Zhenhua Liu, Siwei
  Ma, Chunjing Xu, Chao Xu, and Wen Gao.
\newblock Pre-trained image processing transformer.
\newblock In {\em Proceedings of the IEEE/CVF Conference on Computer Vision and
  Pattern Recognition}, pages 12299--12310, 2021.

\bibitem{chen2205activating}
X Chen, X Wang, J Zhou, and C Dong.
\newblock Activating more pixels in image super-resolution transformer.
\newblock In {\em Proceedings of the IEEE/CVF Conference on Computer Vision and
  Pattern Recognition}, 2023.

\bibitem{mou2022transcl}
Jian~Zhang Chong~Mou.
\newblock Transcl: Transformer makes strong and flexible compressive learning.
\newblock {\em IEEE Transactions on Pattern Analysis and Machine Intelligence
  (TPAMI)}, 2022.

\bibitem{deng2021lau}
Xin Deng, Hao Wang, Mai Xu, Yichen Guo, Yuhang Song, and Li Yang.
\newblock Lau-net: Latitude adaptive upscaling network for omnidirectional
  image super-resolution.
\newblock In {\em Proceedings of the IEEE/CVF Conference on Computer Vision and
  Pattern Recognition}, pages 9189--9198, 2021.

\bibitem{dong2015image}
Chao Dong, Chen~Change Loy, Kaiming He, and Xiaoou Tang.
\newblock Image super-resolution using deep convolutional networks.
\newblock {\em IEEE transactions on pattern analysis and machine intelligence},
  38(2):295--307, 2015.

\bibitem{fakour2018360}
Vida Fakour-Sevom, Esin Guldogan, and Joni-Kristian K{\"a}m{\"a}r{\"a}inen.
\newblock 360 panorama super-resolution using deep convolutional networks.
\newblock 2018.

\bibitem{gu2021interpreting}
Jinjin Gu and Chao Dong.
\newblock Interpreting super-resolution networks with local attribution maps.
\newblock In {\em Proceedings of the IEEE/CVF Conference on Computer Vision and
  Pattern Recognition}, pages 9199--9208, 2021.

\bibitem{hu2023DEARGAN}
Yujie Hu, Yinhuai Wang, and Jian Zhang.
\newblock Dear-gan: Degradation-aware face restoration with gan prior.
\newblock {\em IEEE Transactions on Circuits and Systems for Video Technology
  (TCSVT)}, 2023.

\bibitem{ledig2017photo}
Christian Ledig, Lucas Theis, Ferenc Husz{\'a}r, Jose Caballero, Andrew
  Cunningham, Alejandro Acosta, Andrew Aitken, Alykhan Tejani, Johannes Totz,
  Zehan Wang, et~al.
\newblock Photo-realistic single image super-resolution using a generative
  adversarial network.
\newblock In {\em Proceedings of the IEEE conference on computer vision and
  pattern recognition}, pages 4681--4690, 2017.

\bibitem{busifusion}
Jiabao Li, Yuqi Li, Chong Wang, Xulun Ye, and Wolfgang Heidrich.
\newblock Busifusion: Blind unsupervised single image fusion of hyperspectral
  and rgb images.
\newblock {\em IEEE Transactions on Computational Imaging}, 9:94--105, 2023.

\bibitem{li2022d3c2}
Weiqi Li, Bin Chen, and Jian Zhang.
\newblock D3c2-net: Dual-domain deep convolutional coding network for
  compressive sensing.
\newblock {\em arXiv preprint arXiv:2207.13560}, 2022.

\bibitem{li2021efficient}
Wenbo Li, Xin Lu, Jiangbo Lu, Xiangyu Zhang, and Jiaya Jia.
\newblock On efficient transformer and image pre-training for low-level vision.
\newblock {\em arXiv preprint arXiv:2112.10175}, 2021.

\bibitem{liang2021swinir}
Jingyun Liang, Jiezhang Cao, Guolei Sun, Kai Zhang, Luc Van~Gool, and Radu
  Timofte.
\newblock Swinir: Image restoration using swin transformer.
\newblock In {\em Proceedings of the IEEE/CVF international conference on
  computer vision}, pages 1833--1844, 2021.

\bibitem{lim2017enhanced}
Bee Lim, Sanghyun Son, Heewon Kim, Seungjun Nah, and Kyoung Mu~Lee.
\newblock Enhanced deep residual networks for single image super-resolution.
\newblock In {\em Proceedings of the IEEE conference on computer vision and
  pattern recognition workshops}, pages 136--144, 2017.

\bibitem{mei2021image}
Yiqun Mei, Yuchen Fan, and Yuqian Zhou.
\newblock Image super-resolution with non-local sparse attention.
\newblock In {\em Proceedings of the IEEE/CVF Conference on Computer Vision and
  Pattern Recognition}, pages 3517--3526, 2021.

\bibitem{mou2022metric}
Chong Mou, Yanze Wu, Xintao Wang, Chao Dong, Jian Zhang, and Ying Shan.
\newblock Metric learning based interactive modulation for real-world
  super-resolution.
\newblock In {\em Computer Vision--ECCV 2022: 17th European Conference, Tel
  Aviv, Israel, October 23--27, 2022, Proceedings, Part XVII}, pages 723--740.
  Springer, 2022.

\bibitem{shi2016real}
Wenzhe Shi, Jose Caballero, Ferenc Husz{\'a}r, Johannes Totz, Andrew~P Aitken,
  Rob Bishop, Daniel Rueckert, and Zehan Wang.
\newblock Real-time single image and video super-resolution using an efficient
  sub-pixel convolutional neural network.
\newblock In {\em Proceedings of the IEEE conference on computer vision and
  pattern recognition}, pages 1874--1883, 2016.

\bibitem{sun2019distilling}
Xiaopeng Sun, Wen Lu, Rui Wang, and Furui Bai.
\newblock Distilling with residual network for single image super resolution.
\newblock In {\em 2019 IEEE International Conference on Multimedia and Expo
  (ICME)}, pages 1180--1185. IEEE, 2019.

\bibitem{sun2017weighted}
Yule Sun, Ang Lu, and Lu Yu.
\newblock Weighted-to-spherically-uniform quality evaluation for
  omnidirectional video.
\newblock {\em IEEE signal processing letters}, 24(9):1408--1412, 2017.

\bibitem{vaswani2017attention}
Ashish Vaswani, Noam Shazeer, Niki Parmar, Jakob Uszkoreit, Llion Jones,
  Aidan~N Gomez, {\L}ukasz Kaiser, and Illia Polosukhin.
\newblock Attention is all you need.
\newblock {\em Advances in neural information processing systems}, 30, 2017.

\bibitem{wang2021real}
Xintao Wang, Liangbin Xie, Chao Dong, and Ying Shan.
\newblock Real-esrgan: Training real-world blind super-resolution with pure
  synthetic data.
\newblock In {\em Proceedings of the IEEE/CVF International Conference on
  Computer Vision}, pages 1905--1914, 2021.

\bibitem{wang2018esrgan}
Xintao Wang, Ke Yu, Shixiang Wu, Jinjin Gu, Yihao Liu, Chao Dong, Yu Qiao, and
  Chen Change~Loy.
\newblock Esrgan: Enhanced super-resolution generative adversarial networks.
\newblock In {\em Proceedings of the European conference on computer vision
  (ECCV) workshops}, pages 0--0, 2018.

\bibitem{wang2023gpsr}
Yinhuai Wang, Yujie Hu, Jiwen Yu, and Jian Zhang.
\newblock Gan prior based null-space learning for consistent super-resolution.
\newblock In {\em AAAI Conference on Artificial Intelligence (AAAI)}, 2023.

\bibitem{wang2023ddnm}
Yinhuai Wang, Jiwen Yu, and Jian Zhang.
\newblock Zero-shot image restoration using denoising diffusion null-space
  model.
\newblock In {\em International Conference on Learning Representations (ICLR)},
  2023.

\bibitem{animeSR}
Yanze Wu, Xintao Wang, Gen Li, and Ying Shan.
\newblock Animesr: Learning real-world super-resolution models for animation
  videos.
\newblock In {\em Advances in Neural Information Processing Systems}, 2022.

\bibitem{yoon2022spheresr}
Youngho Yoon, Inchul Chung, Lin Wang, and Kuk-Jin Yoon.
\newblock Spheresr: 360deg image super-resolution with arbitrary projection via
  continuous spherical image representation.
\newblock In {\em Proceedings of the IEEE/CVF Conference on Computer Vision and
  Pattern Recognition}, pages 5677--5686, 2022.

\bibitem{yu2023osrt}
Fanghua Yu, Xintao Wang, Mingdeng Cao, Gen Li, Ying Shan, and Chao Dong.
\newblock Osrt: Omnidirectional image super-resolution with distortion-aware
  transformer.
\newblock In {\em Proceedings of the IEEE/CVF Conference on Computer Vision and
  Pattern Recognition}, 2023.

\bibitem{zhang2019ranksrgan}
Wenlong Zhang, Yihao Liu, Chao Dong, and Yu Qiao.
\newblock Ranksrgan: Generative adversarial networks with ranker for image
  super-resolution.
\newblock In {\em Proceedings of the IEEE/CVF International Conference on
  Computer Vision}, pages 3096--3105, 2019.

\bibitem{zhang2018image}
Yulun Zhang, Kunpeng Li, Kai Li, Lichen Wang, Bineng Zhong, and Yun Fu.
\newblock Image super-resolution using very deep residual channel attention
  networks.
\newblock In {\em Proceedings of the European conference on computer vision
  (ECCV)}, pages 286--301, 2018.

\bibitem{zhang2018residual}
Yulun Zhang, Yapeng Tian, Yu Kong, Bineng Zhong, and Yun Fu.
\newblock Residual dense network for image super-resolution.
\newblock In {\em Proceedings of the IEEE conference on computer vision and
  pattern recognition}, pages 2472--2481, 2018.

\bibitem{zhang2020toward}
Yupeng Zhang, Hengzhi Zhang, Daojing Li, Liyan Liu, Hong Yi, Wei Wang, Hiroshi
  Suitoh, and Makoto Odamaki.
\newblock Toward real-world panoramic image enhancement.
\newblock In {\em Proceedings of the IEEE/CVF Conference on Computer Vision and
  Pattern Recognition Workshops}, pages 628--629, 2020.

\bibitem{zhou2022spatial}
Man Zhou, Jie Huang, Keyu Yan, Hu Yu, Xueyang Fu, Aiping Liu, Xian Wei, and
  Feng Zhao.
\newblock Spatial-frequency domain information integration for pan-sharpening.
\newblock In {\em Computer Vision--ECCV 2022: 17th European Conference, Tel
  Aviv, Israel, October 23--27, 2022, Proceedings, Part XVIII}, pages 274--291.
  Springer, 2022.

\bibitem{DFUS}
Man Zhou, Hu Yu, Jie Huang, Feng Zhao, Jinwei Gu, Chen~Change Loy, Deyu Meng,
  and Chongyi Li.
\newblock Deep fourier up-sampling.
\newblock {\em arXiv preprint arXiv:2210.05171}, 2022.

\end{thebibliography}
}

\end{document}